\documentclass[useAMS,usenatbib]{mn2e}
\usepackage{graphicx}
\usepackage[figuresright]{rotating}
\usepackage{subfigure}
\usepackage{longtable}
\usepackage{txfonts}

\newcommand{\ii}{\'\i}
\newcommand\ion[2]{#1$\,${\scshape{#2}}}
\newcommand\ohlog{\rm \log(O/H)}

\title[An observational study of the interacting galaxy pair AM\,2322-821]
{The effects of the interaction on the kinematics, stellar population and
metallicity of AM\,2322-821 with Gemini/GMOS}

\author[Krabbe et al.]
{A.~C.~Krabbe$^{1}$\thanks{E-mail:angela.krabbe@gmail.com}, 
M.~G.~Pastoriza$^2$, Cl\'audia~Winge$^3$, I.~Rodrigues$^1$, O.~L.~Dors$^1$
and
\newauthor
D.~L.~Ferreiro$^4$\\
$^1$ Universidade do Vale do Para\'iba, Av. Shishima Hifumi, 2911, Cep
12244-000,
S\~ao Jos\'e dos Campos, SP, Brazil\\
$^2$ Instituto de F\ii sica, Universidade Federal 
do Rio Grande do Sul, Av.~Bento Gon\c{c}alves, 9500, 
Cep 91359-050, Porto Alegre, RS, Brazil\\
$^3$ Gemini Observatory, c/o AURA Inc., Casilla 603, La Serena, Chile\\
$^5$ IATE, Observat\'orio Astron\'omico, Universidad Nacional de 
C\'ordoba, Laprida 854, 5000, C\'ordoba, Argentina}

\begin{document}

\date{Accepted- 2011 April 19. Received- 2010 July 9.}

\pagerange{\pageref{firstpage}--\pageref{lastpage}} \pubyear{2010}

\maketitle

\label{firstpage}

\begin{abstract}
We present an observational study about the impacts of the interactions
in the kinematics, stellar populations, and oxygen abundances of the components 
of the galaxy pair AM\,2322-821. A fairly symmetric rotation curve for the
companion (AM\,2322B) galaxy with a deprojected velocity amplitude of  110 km
s$^{-1}$ was obtained,  and a dynamical mass
of $ 1.1 - 1.3 \times 10^{10} M_{\sun}$ within  a radius of 4 kpc was  estimated
using 
this deprojected velocity. Asymmetries in the radial velocity field were
detected for the companion, very likely due the interaction between the
galaxies. The interaction between the main and companion galaxies was modelled
using numerical 
N-body/hydrodynamical simulations, with  the result indicating that the current
stage of the system  would be about 90 Myr after 
perigalacticum. The spatial variation in the distribution of the
stellar-population components in both galaxies was
analysed using the stellar population synthesis code {\sc STARLIGHT}.
The companion galaxy is  dominated by a very young (t $ \leq 1\times10^{8}$ yr)
population, with the fraction of this population  
to the total flux  at $\lambda\, 5\,870\, \AA$, increasing outwards in the
galaxy disc.  
On the other hand, the stellar population of AM\,2322A
is heterogeneous along  the slit positions observed.
Spatial profiles of    oxygen abundance in the gaseous phase were obtained using
two diagnostic diagrams ( $R_{23}$=([\ion{O}{ii}]$\lambda\,3727+$[\ion{O}{iii}]
$\lambda\,4959+$[\ion{O}{iii}]  $\lambda\,5007)/$H$\beta$ vs.
[\ion{O}{iii}]$\lambda\,5007$/[\ion{O}{ii}]$\lambda\,3727$ and
[\ion{O}{iii}]$\lambda\,5007$/[\ion{O}{ii}]$\lambda\,3727$
vs.[\ion{N}{ii}]$\lambda\,6584$/[\ion{O}{ii}]$\lambda\,3727$), where we compared
the observed values with the ones obtained from photoionization models. 
  Such gradients of oxygen abundance are significantly 
flatter for this pair of galaxies than
in typical isolated spiral galaxies. This metallicity distribution is interpreted as the gradients having been destroyed by interaction-induced
gas flows from the outer parts to the centre of the galaxy.
\end{abstract}


\section{Introduction}
Galaxy interactions and merger events play an important role on the evolution
and the stellar formation history of galaxies.
Interacting/merging galaxies show enhanced star formation when compared with
isolated objects, as indicated by
different studies \citep{kennicutt87, sekiguchi92,donzelli97, barton03,geller06,
woods07}.
This enhancement  has been observed as being a function of the projected galaxy
pair separation
(e.g. \citealt{barton00, lambas03, nikolic04}), as well as being
stronger in low-mass  than in high-mass galaxies (e.g.
\citealt{woods07,ellison08}).  

The induced star formation associated with the gas motions created by the
interaction also is
expected to have an impact in the chemical state of the galaxies. 
As star formation enriches the interstellar medium via nucleosynthesis, inflows
of metal poor gas from the outer parts of the galaxy can decrease the
metallicity in inner regions
and  modify the radial abundance gradients across spiral discs. In fact, 
studies have found that 
interacting galaxies do not follow the well established correlation between
luminosity and  metallicity found in normal disc galaxies. The central regions of these 
galaxies are underabundant when compared to isolated galaxies of similar mass 
\citep{kewley06,rupke08,ellison08,michel-dansac08,peeples09}.

Shallower metallicity gradients have been found in barred galaxies  
and  explained by the action of inward and outward radial flows of interstellar
gas induced by the non-axisymmetric potential of bars (e.g. \citealt{friendli94,sellwood93,roy97}).
N-body/Smoothed Particle Hydrodynamics numerical simulations of equal-mass
mergers, although ignore
the presence of ongoing star formation, predict that the radial metallicity
gradients in disc galaxies flatten shortly after the first pericentre passage,
due to the radial mixing of gas \citep{rupke2010}.
 Recently, 
\citet{kewley10} 
determined the metallicity gradients for 8 galaxy pairs  with mass ratio near unity and showed that they were significantly shallower
than  those in isolated spiral galaxies. Futhermore, 
\citet{krabbe08} found for the  interacting pair  AM\,2306-721  with  a mass ratio of 2:1 that
the disc of the main galaxy showed a clear radial metallicity gradient, while the secondary presented 
a relatively homogeneous oxygen abundance. These authors interpreted the absence of abundance gradient in the secondary galaxy in terms of mixing the low metallicity gas from the outer parts with the rich metal gas of the centre of the galaxy. Is this picture always reproduced in others minor mergers? This question was not adressed before  by theoretical merger simulations and observations, except the above study 
of AM\,2306-721. Therefore, more observational studies 
of mergers  with different mass ratios should provide useful insights to answer the above  question. We have selected from \citet{ferreiro04} several systems to study the effects of the kinematics, stellar population and gradient abundances of the galaxies in minor mergers, 
where the first results of this programme were presented for 
AM\,2306-721 \citep{krabbe08}.


This paper presents the results  for the system  AM\,2322-821, which  is morphologically very similar to AM\,2306-721, but 
its mass ratio is much lower than the latter.
AM\,2322-821 is composed of a SA(r)c galaxy with
disturbed arms (hereafter, AM\,2322A) in interaction with an irregular galaxy
(hereafter, AM\,2322B). 
Both galaxies contain very luminous  \ion{H}{ii} regions with H$\alpha$
luminosity 
in the range of $2.53\times\,10^{39}\,<\,$L(H$\alpha\,)\,
<\,1.45\times\,10^{41}$erg\, s$^{-1}$ 
as estimated from H$\alpha\,$ images   
and high star formation rate in the range of 0.02 to 1.15  $ M_{\sun}$/yr
\citep{ferreiro08}.



The present paper is organized as follows: in Section \ref{datared},  we
summarize the observations and data reduction. 
The gas kinematics of each galaxy and the numerical
N-body/hydrodynamical simulations of the interaction are presented in 
Sections \ref{vel} and \ref{numsim}, respectively. In Section
\ref{sintese}, we present the stellar population synthesis. The
metallicity analysis is in Section \ref{emission}, and the
conclusions are summarized in 
Section \ref{final}.

\section{Observations and data reduction} \label{datared}
Long slit spectroscopic data were obtained on 29/30
June 2006, 01/02 July 2006, and 27/28 July 2008 with the Gemini Multi-Object
Spectrograph at Gemini South, as part of poor weather programmes
GS-2006A-DD-6 and GS-2008A-Q-206.   Spectra in the range 3\,450 to
7\,130\AA\ were acquired with  two settings with the B600 grating, and the 1$\arcsec$
slit, keeping  a compromise between
spectral resolution ($\thicksim5.5\,\AA$), spectral coverage and slit losses (due to the Image Quality = ANY constraint).
 The blue setting provided a wavelength coverage of 3\,450 to 6\,280\AA\ and the red setting of 4\,280 to 7\,130\AA\ at about the same 
spectral resolution.
The frames were binned on-chip by 4 and 2 pixels
in the spatial and spectra directions, respectively, resulting in a
spatial scale of 0.288 $\arcsec$\,pxl$^{-1}$, and 0.9\AA\,pxl$^{-1}$
dispersion.

\begin{figure}
\centering
\includegraphics*[width=\columnwidth]{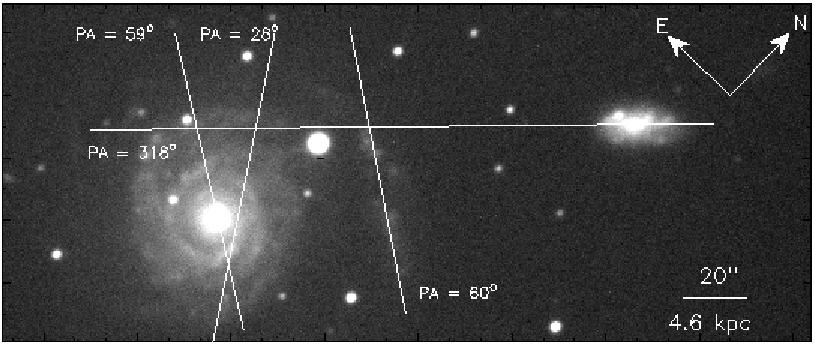}  				
    
\caption{GMOS-S $r'$-band image of AM\,2322-821 with the observed slit
positions.}
\label{field}
\end{figure}

Spectra were taken at four different position angles on the sky, 
  with the goal of observing the nucleus and the brigthest 
regions of the galaxies, as well as, one spiral arm that 
is away from the main galaxy.
PA=59$\degr$ is in the slit position crossing the nucleus of AM\,2322A; 
PA=28$\degr$ is cutting across main body of primary, but not across the nucleus (offset of about 8$\arcsec$ NW from nucleus); 
the slit position at PA=60$\degr$ is located off the disc of the main component,
along the NW spiral arm 
(located between the main and secondary component). The PA=318$\degr$ slit
position is oriented along 
the main axis disc of the secondary component (AM\,2322B) and also along the AM
2322A NE spiral arm.   The PA=60$\degr$ was observed only in the red spectra, and thus in this slit,  
it was  possible only to study  the ionized gas kinematics, but not their stellar population and the gas phase O/H abundance.
Figure 
\ref{field} shows the four slit positions on the  GMOS-S $r'$ acquisition image.

The exposure time on each single frame was limited to 700 seconds to 
minimize the effects of cosmic rays, with multiple frames being obtained
for each slit position to achieve a suitable signal. The slit positions are
shown in Fig. \ref{field}, 
superimposed on the $r'$-band image of the pair. Table \ref{observ}
gives the journal of observations. Conditions during both runs were
not photometric, with thin cirrus and image quality in the range 0.5 $\arcsec$
to 2.0 $\arcsec$ (as measured from stars in the acquisition images taken
just prior to the spectroscopic observations).

The spectroscopic data reduction was carried out  using 
the {\sc gemini.gmos} package as well as  generic {\sc IRAF} tasks.
  We followed
the standard procedure: (1) the data were bias subtracted and flatfielded; 
(2) the wavelengh calibration was established from the Cu-Ar arc frames with  typical residuals of 0.2 \AA\,
and applied to the object frames; (3)  the individual spectra of same slit positions and wavelengh range were averaged
with cosmic ray rejection; (4) the object frames were sky subtracted interactively using the {\sc gsskysub} task, 
which use a background sample of off-object 
areas to fit a function to the  specified rows, and this fit is 
then subtracted from the column of each spectra; (5) the spectra were relative flux calibrated using observations of a flux
standard star taken with the same set up as the science observations; (6) finally, one-dimensional spectra were 
extracted  from the two-dimensional spectra by  summing over six rows along 
the spatial direction. Each spectrum therefore comprises the flux contained in an aperture of 1$\arcsec \times
1.73 \arcsec$. Assuming a distance of 49.6 Mpc for AM\,2322-821 system
barycenter (estimated from the radial velocities derived in Section \ref{vel}
and total masses of AM\,2322A and AM\,2322B, estimated in Section
\ref{numsim}, and adopting $H_0$=75\,km\,s$^{-1}$\,Mpc$^{-1}$), this
aperture corresponds to a region of 241$\times$419\,pc$^2$ for
AM\,2322-821. The nominal centre of each
slit was chosen to be the continuum peak at $\lambda\,5735\AA$. 
The mismatch between the blue and red spectra for all the apertures extracted was lower than about
5 \% over the spectral range where both spectra overlapped. We  decided  not to co-add them.

\begin{table}
\caption{Journal of observations}
\label{observ}
\begin{tabular}{lccc}
\noalign{\smallskip}
\hline
\noalign{\smallskip}
Date (UT) & Exposure time(s) & PA (\degr)& $\Delta \lambda (\AA)$ \\
\noalign{\smallskip}
\hline
\noalign{\smallskip}
2006/06/29  &  3$\times$600    & 318	 & 4280-7130   \\
2006/06/30  &  6$\times$600    & 60	          & 4280-7130   \\
2006/07/01  &  2$\times$600    & 28	          & 4280-7130  \\
2006/07/01  &  3$\times$700    & 59	          & 4280-7130    \\
2006/07/01  &  3$\times$600    & 318	 & 4280-7130    \\
2008/07/27  &  6$\times$600    & 28	          & 3450-6270     \\
2008/07/27  &  6$\times$600    & 59	          & 3450-6270     \\
2008/07/27  &  6$\times$600    & 318	 & 3450-6270   \\
\noalign{\smallskip}
\hline
\noalign{\smallskip}
\end{tabular}
\end{table}

\section{Ionized gas kinematics}
\label{vel}
The radial velocity was estimated from the strongest emission lines present in
the spectra, namely 
$\rm H\beta$ $\lambda 4861$,[\ion{O}{iii}] $\lambda 5007$, 
$\rm H\alpha$ $\lambda 6563$ and  [\ion{N}{ii}] $\lambda 6584$.
The final radial velocity for each spectrum was obtained by averaging
the individual measurements from the detected emission lines, and the errors
were estimated from the
standard deviation of the individual measurements around the mean.

\begin{figure}
\includegraphics*[angle=-90,width=\columnwidth]{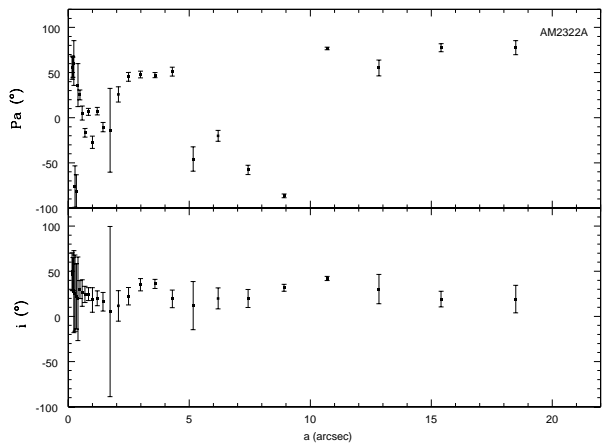}			
	     
\caption{Results of the isophote fitting as a function of the distance along the
major axis a in arcseconds for AM\,2322A. 
Top panel: inclination of the galaxy. Bottom panel: position angle of the major
axis.}
\label{isofota_main}
\end{figure}

\begin{figure}
\includegraphics*[angle=-90,width=\columnwidth]{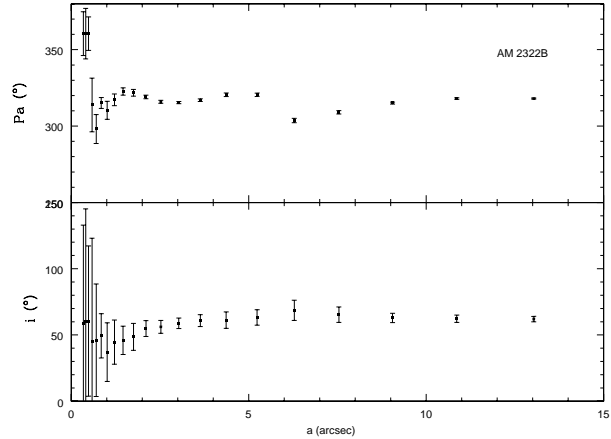}			
	     
\caption{Same as \ref{isofota_main}, but for AM\,2322B.}
\label{isofota}
\end{figure}

The inclination of each  galaxy  with respect to the
plane of the sky was  computed as $\cos(i)=b/a$, where $a$ and $b$ is
the minor and major semi-axes of the galaxy, respectively. The 
minor and major semi-axes as well as, the 
position angle of the major axis of each galaxy were obtained  
from the acquisition images in the $r'$ filter, using a simple
isophotal fitting with the {\sc IRAF stsdas.ellipse} task. The fitting results
for the position angle of the major axis and the inclination 
as a function of the projected distance in arcseconds along the isophotal major
axis are shown in Figs \ref{isofota_main} and 
\ref{isofota}, for AM\,2322A and AM\,2322B, respectively.

For AM\,2322B, the position angle of the major axis and the inclination of the
galaxy are nearly  constant out to a 
$5\arcsec$ radius. The  resulting values for the disc inclination and position
angle of the line of nodes are $i=63\degr$ 
and $\psi_0= 318\degr$, respectively. For AM\,2322A, the position angle of the
major axis and the inclination of the galaxy
as measured from the isophotal fitting show significant variations with radius.
According to \citet{bender87}, the 
interaction among galaxies can  cause isophotal twisting, so these variations
can be due to the disturbed
morphology of the galaxy, 
added to the fact that the GMOS r'-band image is not deep and star forming
regions and spiral structure are dominating in these wavelengths.
If we considered the  external isophotes of AM\,2322A,
the values found are about $\psi_0= 75\degr $ and  $i=20\degr$.

\citet{ferreiro04} estimated the inclination of the galaxies in this pair
measuring  the major and minor diameters of  the 24 mag arcsec$^{-2}$ isophote from 
images in the $B$ filter. They found $i=44\degr$ and  $i=54\degr$  for AM\,2322A
and AM\,2322B, respectively. 

\begin{figure*}
\subfigure{
\includegraphics*[angle=-90,width=\columnwidth]{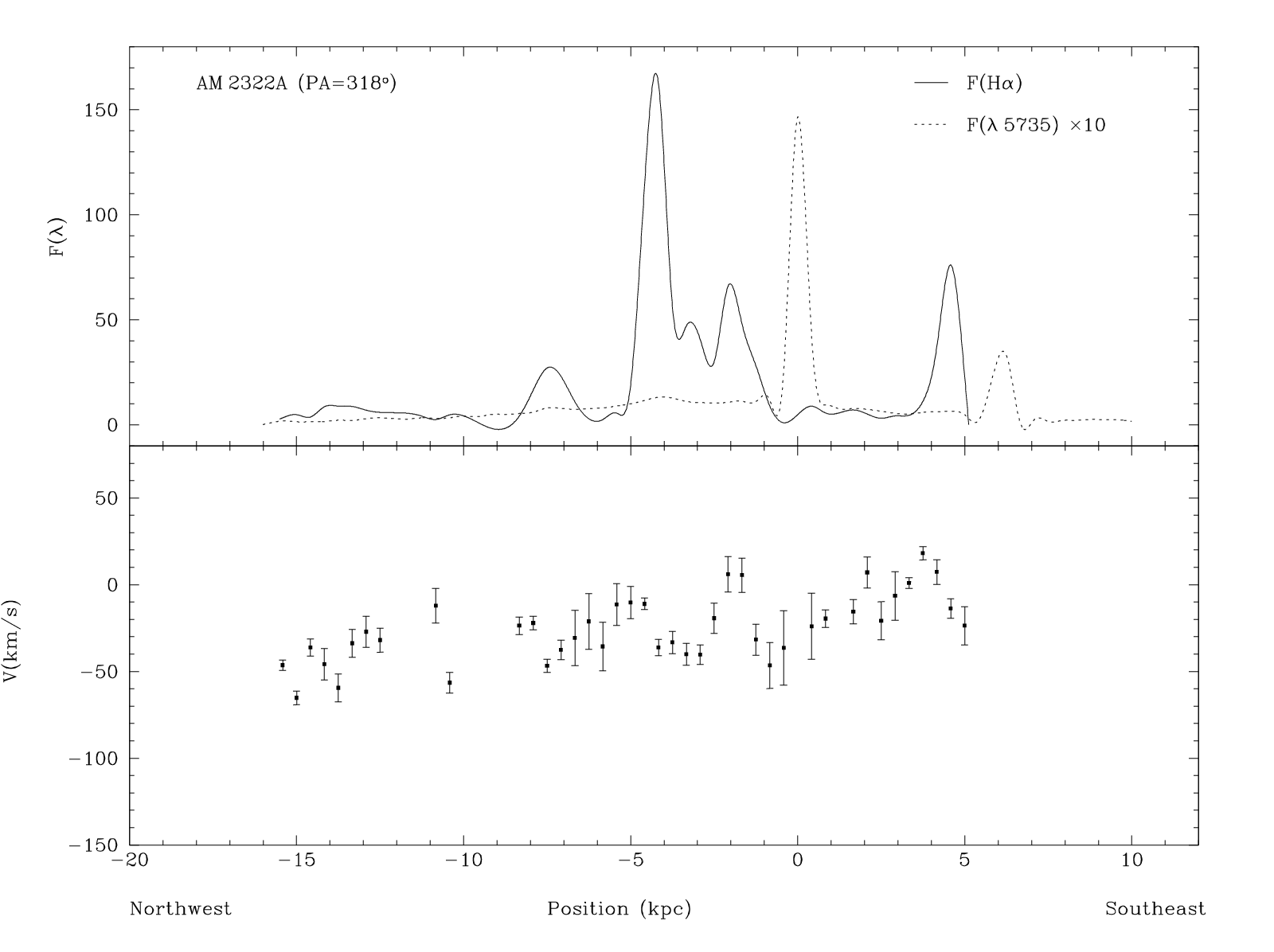}}
\subfigure{
\includegraphics*[angle=-90,width=\columnwidth]{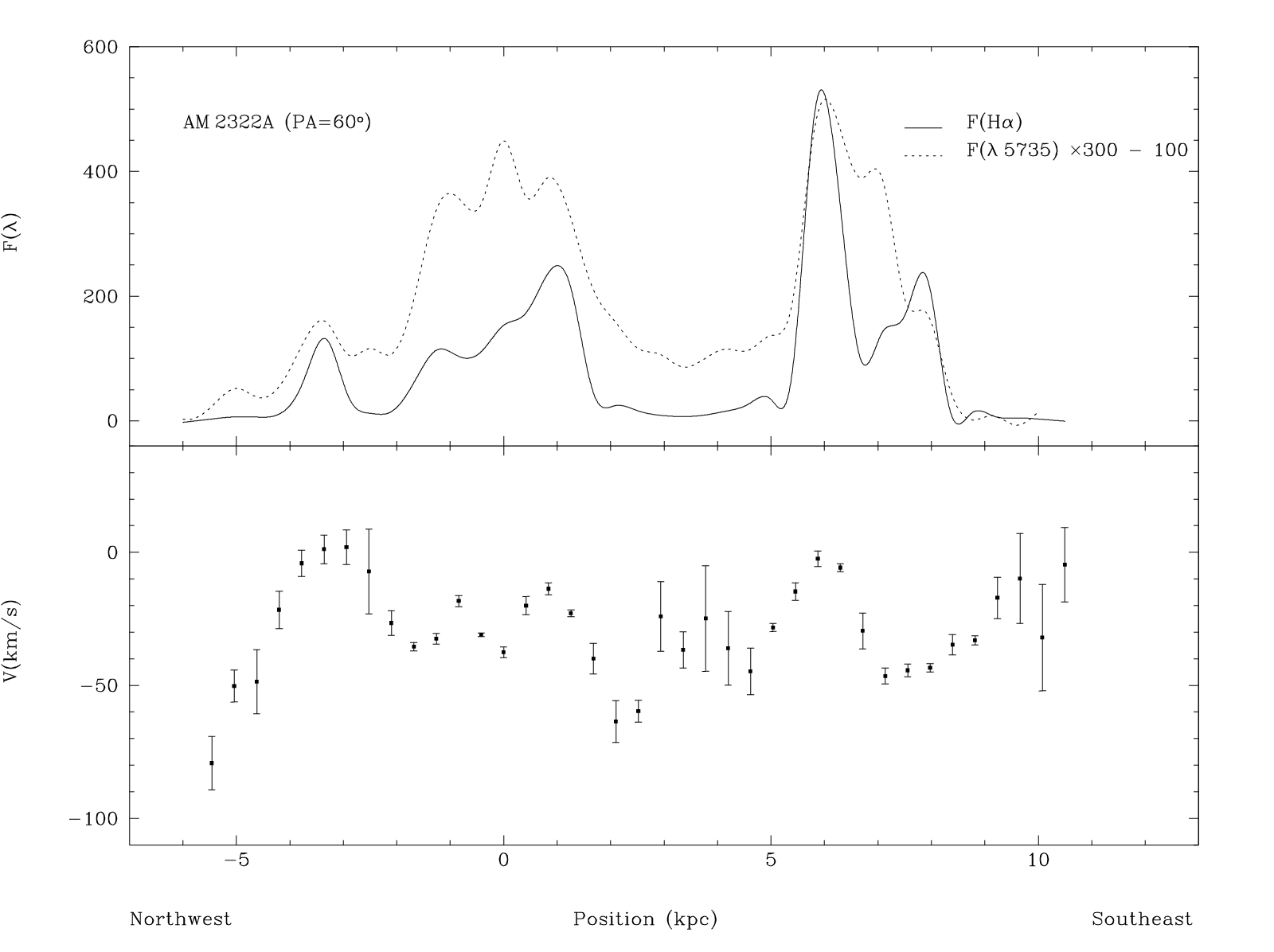}}
\subfigure{
\includegraphics*[angle=-90,width=\columnwidth]{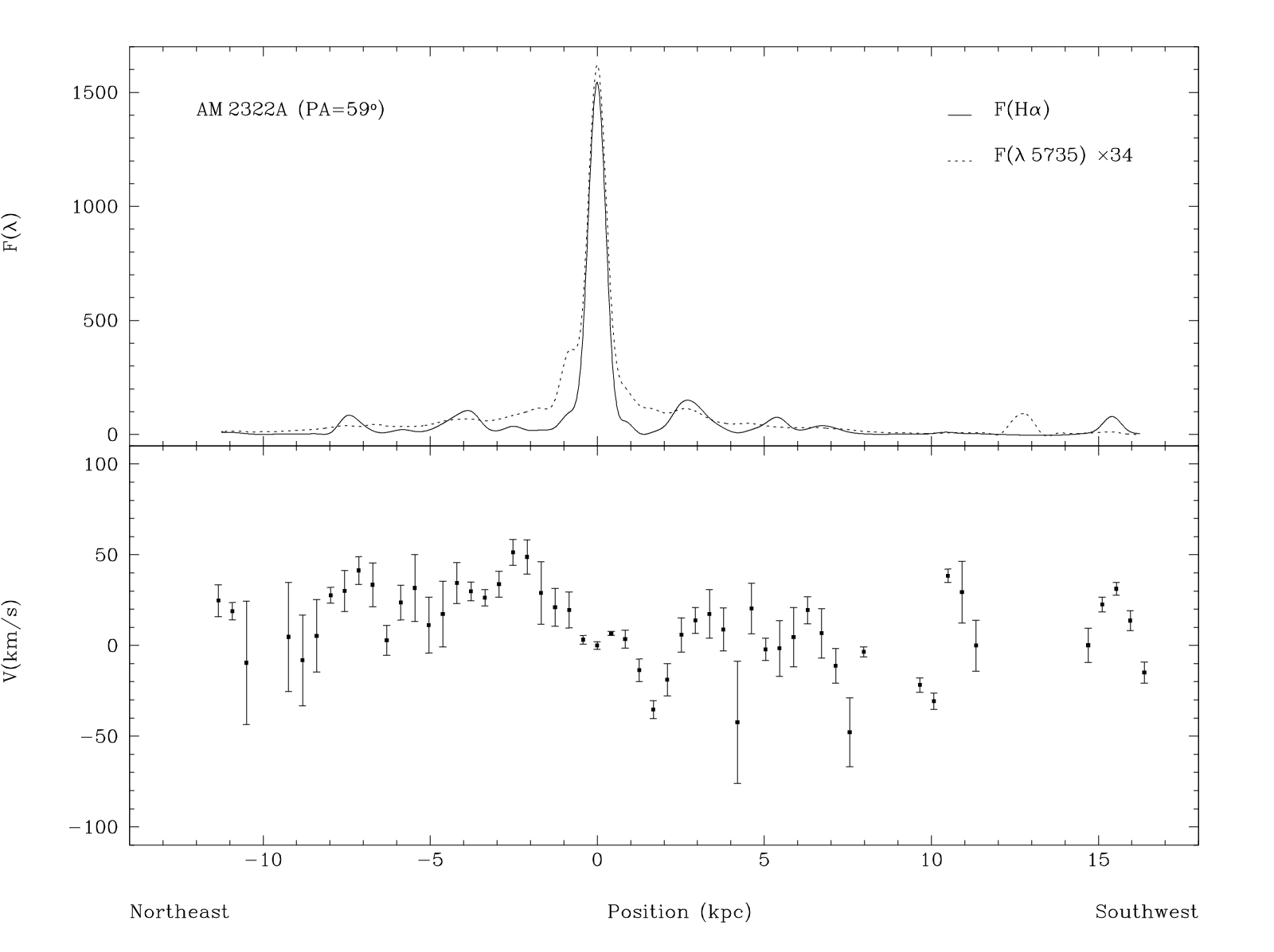}}
\subfigure{
\includegraphics*[angle=-90,width=\columnwidth]{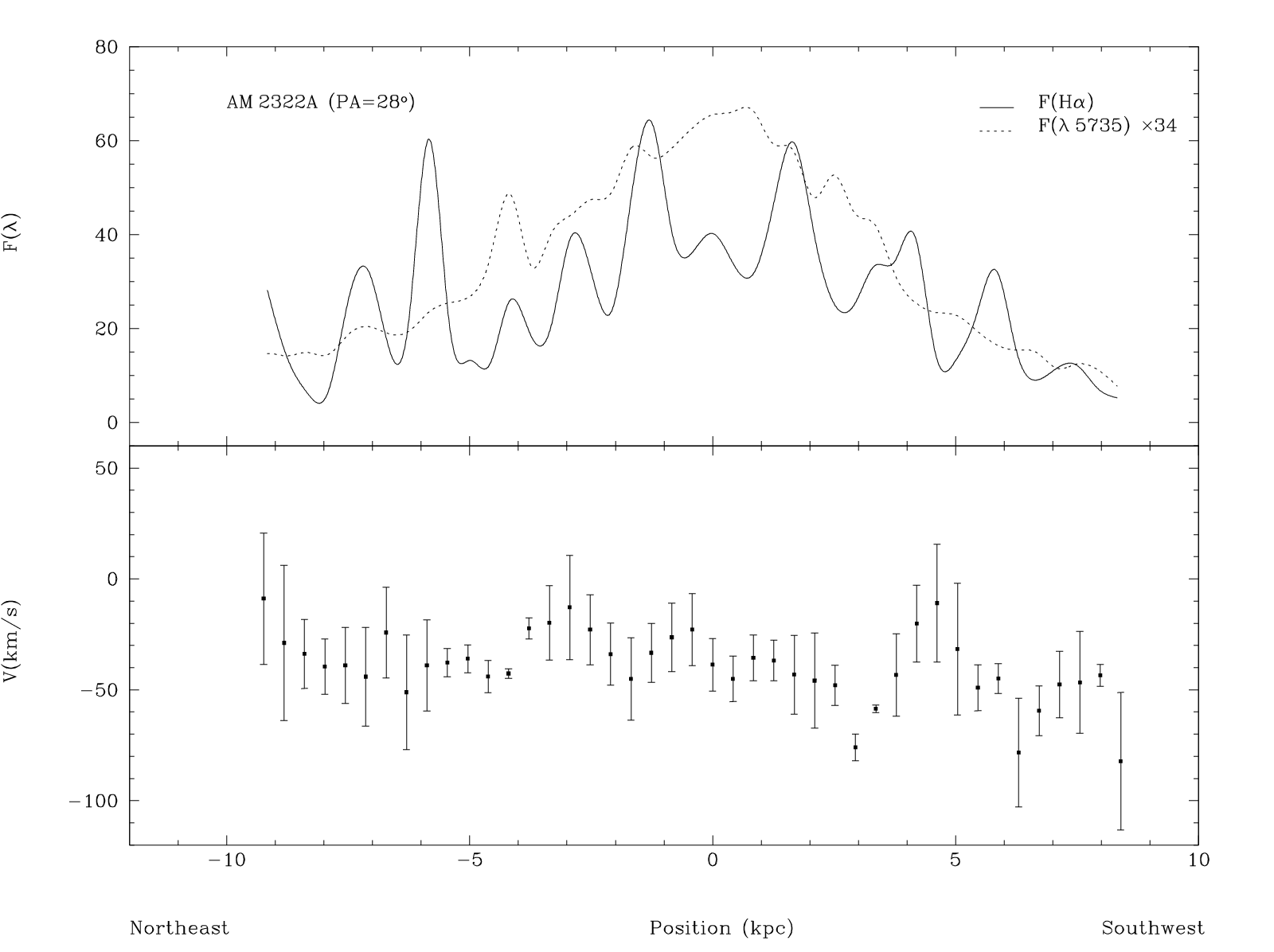}}
\caption{Observed $\rm H\alpha$ and $\lambda\, 5735$ flux (in units of 10$^{17}$
ergs
cm$^{-2}$ s$^{-1}$) and mean radial velocity  as a function of the distance to
the centre of the slit along PA=318\degr\,, PA=60\degr\,, PA=59\degr\, and PA=28\degr\ for AM\,2322A.
The velocity scale correspond to the
observed and model values after subtraction of the systemic velocity
of the galaxy, without correction by the inclination in the plane of
the sky. }
\label{main_velocity}
\end{figure*}

\begin{figure}
\includegraphics*[angle=-90,width=\columnwidth]{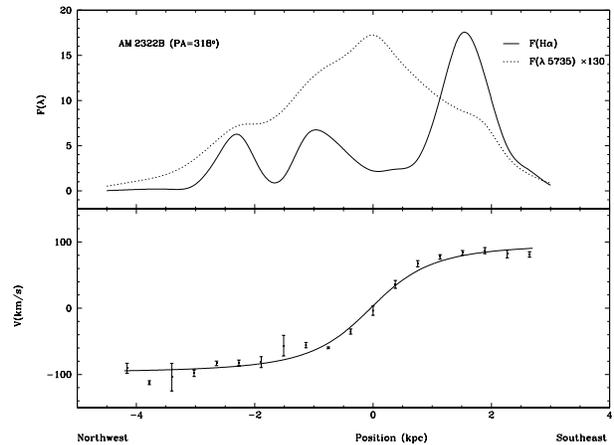}		
		     
\caption{Observed $\rm H\alpha$ and $\lambda\, 5735$ flux (in units of 10$^{17}$
ergs
cm$^{-2}$ s$^{-1}$) and mean radial velocity  as a function of apparent
galactocentric distance
along PA=318\degr\, for AM\,2322B. The velocity scale correspond to the
observed and model values after subtraction of the systemic velocity
of the galaxy, without correction by the inclination in the plane of
the sky.}
\label{sec_velocity}
\end{figure}

The rotation curves and the spatial profiles of the $\rm H\alpha$ emission and
$\lambda\, 5735$ continuum flux along the observed slit positions are presented
in Figs. \ref{main_velocity} and \ref{sec_velocity} for AM\,2322A and AM\,2322B,
respectively.

As can be seen in Fig. \ref{main_velocity}, AM\,2322A   does
not have a 
well defined and symmetric rotation curve along  to the  observed slit
positions, 
indicating that the inclination of the galaxy must be quite low and that this
object is actually  being  seen near face-on. 
Then, the real inclination of the galaxy must be lower than the one obtained
from \citet{ferreiro04} and perhaps even from our own estimation.

The heliocentric velocity  of the main galaxy is taken  to be the radial
velocity
measured at the nominal centre (continuum peak) along the slit position at 
PA=59\degr\,  or $v_r=$ 3\,739 km s$^{-1}$.

AM\,2322B shows a fairly symmetric rotation curve and for this galaxy we adopted
a very simple 
approximation for the observed velocity distribution,  assuming that the gas
moves under a
logarithmic gravitational potential, following circular 
orbits close to a plane $P(i,\psi_{0})$, characterized by its inclination 
to the plane of the sky $(i)$ and the position angle (PA) 
of the line of nodes  $\psi_0$. This assumption results in an observed radial
circular velocity
$v(r,\psi)$ in the plane of the sky given by \citet{bertola91}:

\begin{equation}
v(r,\psi)= V_{s} + \frac{V_{0}R
\cos(\psi-\psi_{0})\sin(i)\cos(i)}{\sqrt{R^{2}\eta + R_{c}^{2}\cos^{2}(i)}},
\label{v_mol}
\end{equation}
with

\begin{equation}
\eta \equiv [\sin^{2}(\psi-\psi_{0}) + \cos^{2}(i)\cos^{2}(\psi-\psi_{0})],
\label{v_molcont}
\end{equation}
where $V_{s}$ is the systemic velocity, $R$ is the radius in the plane of 
the galaxy, and $V_{0}$ and $R_{c}$  are parameters that define 
the amplitude and shape of the curve.  The fit of the rotation curve for this
galaxy is shown in Fig. \ref{sec_velocity}.

The above model for the rotation curve results in an heliocentric velocity of
3\,376 km s$^{-1}$.
The observed radial velocities along the major axis  are well represented by the
model.
The rotation curve is typical of spiral discs, rising shallowly and flattening
at an
observed amplitude of 110 km s$^{-1}$, with small (less than 25 km\,s$^{-1}$)
deviations from  the smooth rotational field, which are commonly observed in
interacting galaxies.

\citet{donzelli97} estimated systemic velocities of 3\,680 and 3\,424
km\,s$^{-1}$ for AM\,2322A and AM\,2322B, respectively, which agree 
within 2 \% with our values.

\begin{table}
\centering
\caption{Kinematical parameters for AM\,2322B}
\label{par}
\begin{tabular}{lccc}
\noalign{\smallskip}
\hline
\noalign{\smallskip}
\noalign{\smallskip}
\noalign{\smallskip}
Parameter & \multicolumn{1}{c}{AM\,2322B (PA=318\degr)}  \\
\noalign{\smallskip}
\hline
\noalign{\smallskip}
$i$ (\degr)     &     63                                  \\
$\psi_0$ (\degr)&     318                                  \\
$V_{s}$ (km/s)  & 3\,376$\pm6$                             \\
$V_{0}$ (km/s)  & 110$\pm6$                              \\
$R_{c}$ (kpc)   & -0.33$\pm$1391                           \\
\noalign{\smallskip}
\hline
\noalign{\smallskip}
\end{tabular}
\end{table}

For AM\,2322B, we can obtain an estimation of the dynamical mass  by assuming
that the mass inside a certain radius is given by 
$ M(R)=RV^{2}/G$. Using the deprojected velocity amplitude of  110 km s$^{-1}$
and a radius of 4 kpc its dynamical mass is $ 1.12 \times 10^{10} M_{\sun}$. It
is important to emphasize that
the maximum radius to which we can observe the gas in emission is
quite certainly smaller than the total radius of the galaxies, so
our estimate of the dynamical mass is a lower limit to the
actual dynamical mass of the system. The 
estimation of the deprojected velocity is also
dependent on the assumed inclination of the galaxies with respect to
the plane of the sky,  so if 
we assume the inclination angle of $i=54\degr$ derived by \citet{ferreiro04},
the dynamical mass would be of 
$ 1.36 \times 10^{10} M_{\sun}$.

\section{Numerical Simulations}
\label{numsim}
Aiming to reconstruct the history of the AM\,2322-821 system and predict its
evolution,
we attempted to  reproduce the interaction between AM\,2322A and AM\,2322B by
running a
series of N-body simulations. The simulations were carried out with the
N-body/SPH code GADGET-2 developed by \citet{2005MNRAS.364.1105S}. Galaxies were
modeled following the prescription of \citet{1993ApJS...86..389H}, where we
included a gaseous disk component. 

\begin{figure}
 \centering
\includegraphics*[width=\columnwidth]{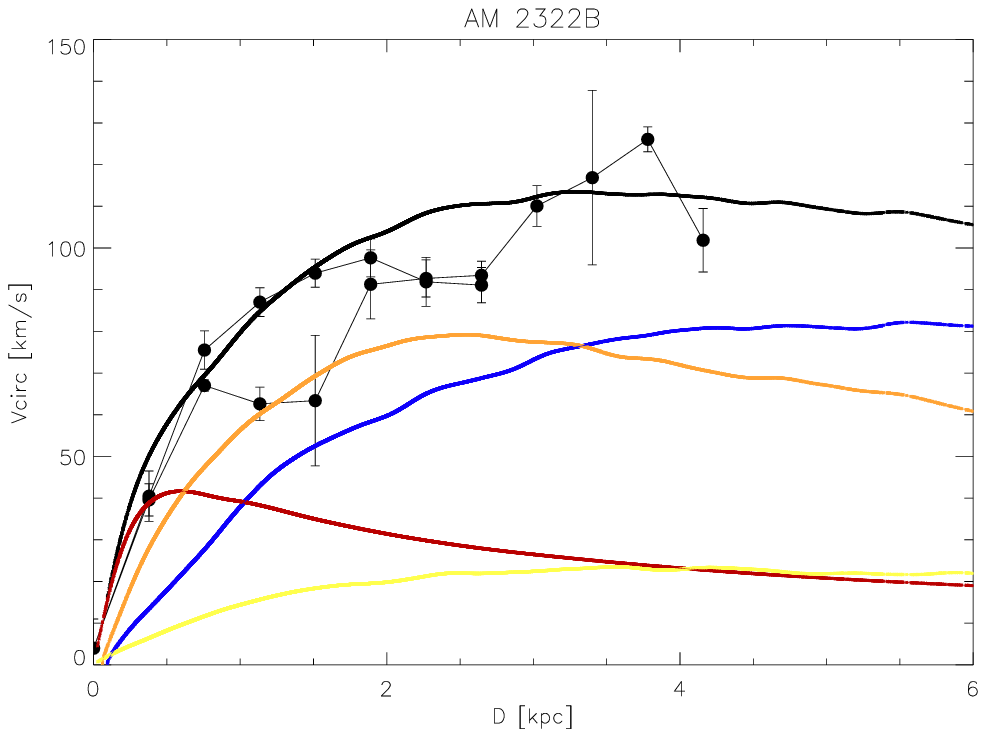}
\caption{Circular velocity curve of the initial model for AM\,2322B (initial
condition for the simulation). Rotation curves of the individual components are
shown: spherical halo in orange, bulge in red, stellar disk in blue and gas disk
in yellow. The total rotation curve is the continuous thick black line. Points
with error bars connected by thin black line segments are the observed radial
velocity values.}
\label{vcirc_obs_model_sec}
\end{figure}

The model parameters for AM\,2322B were constrained from the observed morphology
and rotation curve presented in Sec.~\ref{vel}. The resulting simulated rotation
curve is shown in Fig. \ref{vcirc_obs_model_sec}, where we overplot the observed
circular velocity data. 
\begin{table}
\caption{Parameters used on the simulations}  
\label{tab:modelpars}
\begin{small}
\begin{tabular}{l c c}
\hline
                                & AM\,2322A 	& AM\,2322B \\
\hline
Number of points in disk        & 16384     		&   8192\\
Disk mass               	& 0.6      		&   0.08\\
Disk radial scale length        & 0.7      		&   0.3\\
Disk vertical scale thickness   & 0.14       		&   0.0\\
Reference radius R$_{ref}$      & 1.7        		&   0.8\\
Toomre Q at R$_{ref}$           & 1.5       		&   1.5\\
\hline
Number of points in gas disk    & 16384			&   8192\\
Gas disk mass               	& 0.06       		&   0.01\\
Gas disk radial scale length    & 1.0			&   0.45\\
Gas disk vertical scale thickness	& 0.07      	&   0.01\\
Toomre Q at R$_{ref}$           & 1.2       		&   1.4\\
\hline
Number of points in bulge       & 1024       		&   1024\\
Bulge mass              	& 0.5			&   0.01\\
Bulge radial scale length       & 0.14       		&   0.05\\
\hline
Number of points in spherical halo  & 16384	     	&   8192\\
Halo mass               	& 2.4       		&   0.24\\
Halo cutoff radius          	& 6.0       		&   2.0\\
Halo core radius            	& 0.6        		&   0.4\\
\hline
\smallskip
\end{tabular}
 Notes: Simulations were done in a system of units with
G=1. Model units scales to physical ones such that: length unit is
3.5\,kpc, unit velocity is 262\,km\,s$^{-1}$, unit mass is
$5.586\times 10^{10} \mathrm{M}_\odot$ and unit time is 13.062\,Myr.
\end{small}
\end{table}

In order to build a model for AM\,2322A, we first attempted to measure its
inclination  through isophotal ellipse fitting (see Fig. \ref{isofota_main}),
but the 
results are inconclusive due to the small inclination angle, combined with tidal
distortions 
experienced by the galaxy, which result in strong radial variations on the
resulting isophotal position
angle. The long-slit spectra crossing the disk through two different position
angles (see Figures \ref{field} and \ref{main_velocity}), and none of those
show clear signs of rotation,  so
the rotation curve cannot really be used
to constrain the mass distribution of the galaxy 
as in the case of AM\,2322B. AM\,2322A is modeled as a face-on galaxy. 
A photometric mass estimate of $1.7\times10^{11} M_{\sun}$ is obtained 
using an absolute B magnitude M$_B=-20.98$ \citep{ferreiro04} and assuming a 
mass-to-light ratio of $\gamma_B=4.4$\,M$_\odot$\,L$_\odot^{-1}$ from 
\citet{1979ARA&A..17..135F} as valid for a Hubble type SA(r)c galaxy
\citep{1991rc3..book.....D}.

As usual in this kind of approach, to reproduce the dynamical and morphological
state of the AM\,2322-821 system we have
to solve a reverse problem of finding the orbit followed by the
galaxies from their observed properties, and this is not a fully
determined problem, since the observational data do not provide all
the necessary information. Therefore, in order to set up the initial
conditions for the simulations, we calculate orbits that 
satisfy the requirements given by the observed radial velocity
difference, testing different eccentricities, pericenter distances,
and line of sight direction distances. After that, based on the
observed morphology and previous experience, we select a few orbits to
simulate, from which the one that best fits the observed properties is
selected. 

Different galaxy models were tested, and the parameters of the final 
ones are presented in Table~\ref{tab:modelpars}. The AM\,2322A
model has a total mass of $1.7\times10^{11} M_{\sun}$, and the mass of the
AM\,2322B model is $1.6\times10^{10} M_{\sun}$. 
The mass of each individual
component (disk, bulge, halo, and gas) of both models is also given 
in the table. The disk inclination in the models were chosen to match
the observationally derived values (see Section~\ref{vel}:
AM\,2322A is modeled as a face-on galaxy). The determination of AM\,2322B model 
inclination required several test simulations, because the tidal distortions and 
 warping induced by the interaction modifies the projected axial ratio: initial 
conditions with AM\,2322B model inclined by $i=68^\circ$ led to final 
morfologies that do not match with the observed one. The best results were
obtained 
with AM\,2322B initially inclined by $i=80^\circ$. A total of 75776 particles
were used.

\begin{figure*}
	\centering
	\includegraphics[width=\columnwidth]{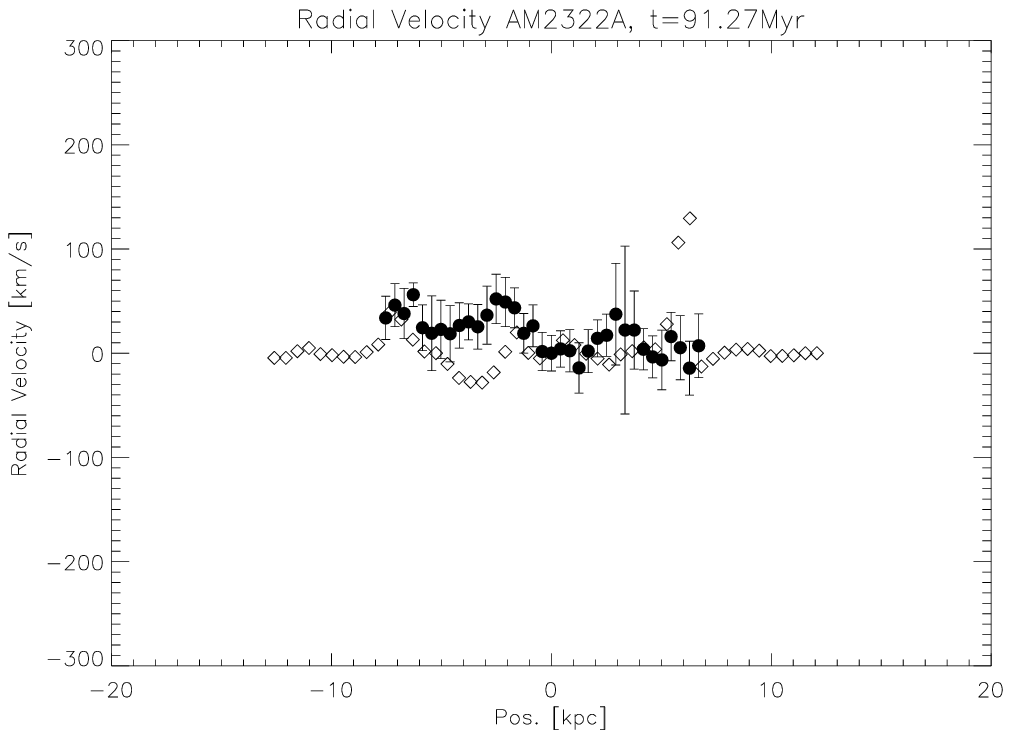}
	\includegraphics[width=\columnwidth]{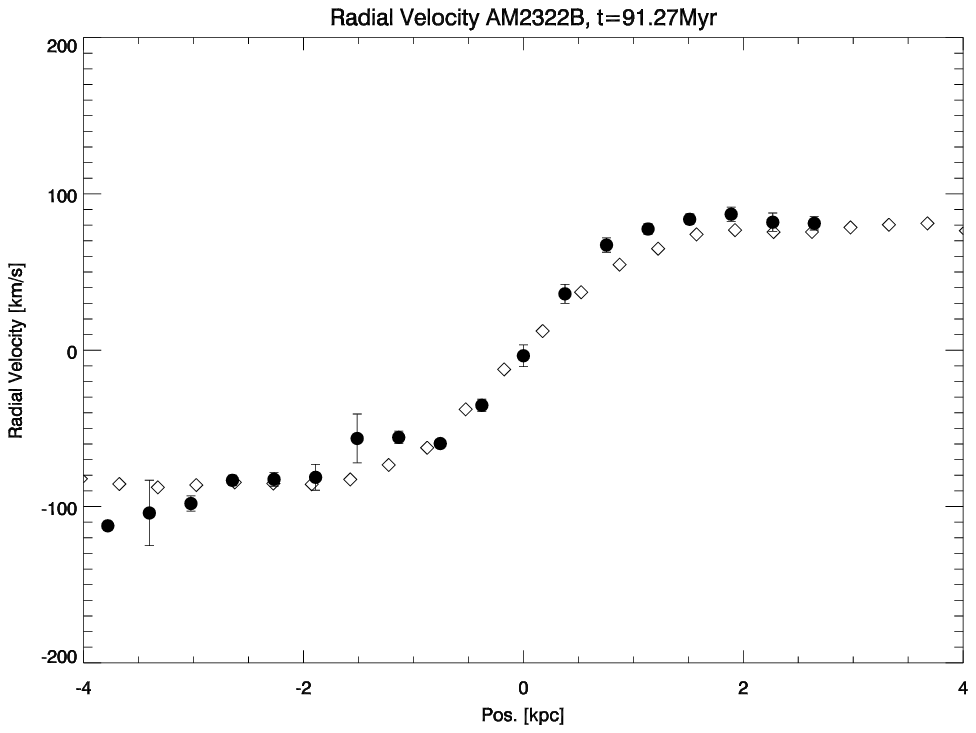}
\caption{Radial velocity curves of the models at the final stage, taken at the
same position angles as the observed ones, for comparison. Left and right panels
are for AM\,2322A and  AM\,2322B, respectively. Losanges are from the
simulations, black dots with error bars are from observations.}
	\label{rotcurve-pos}
\end{figure*}

After several runs, the orbit that best reproduces the observational properties
is a fast hyperbolic orbit, with an eccentricity $e=3.1$ and perigalacticum of
$q=10.5$ kpc. The orbital plane is inclined to the plane of the sky by
$37^\circ$, and intersects the later in a position angle of $80^\circ$.

  The dynamical mass of AM\,2322B, as obtained in
Section~\ref{datared} up to a radius of 4 kpc, amounted to  $ 1.12
\times 10^{10} M_{\sun}$. At the final stage of the simulation, up to
that radius, the model provided  a mass of $ 1.1 \times 10^{10} M_{\sun}$ (the initial
model has a mass of $ 1.22 \times 10^{10} M_{\sun}$ up to that radius). 

The simulation indicates that the spatial distance between both galaxies is
43.8\,kpc (sky projected distance is 35\,kpc). AM\,2322A is closer to us than
AM\,2322B.

Radial velocity curves of model galaxies AM\,2322A and AM\,2322B are presented
in Figure~\ref{rotcurve-pos}. They are compared with the observed radial
velocity data and show that the kinematics of the models is correct.

\begin{figure*}
	\centering
	\fbox{\includegraphics[width=\columnwidth]{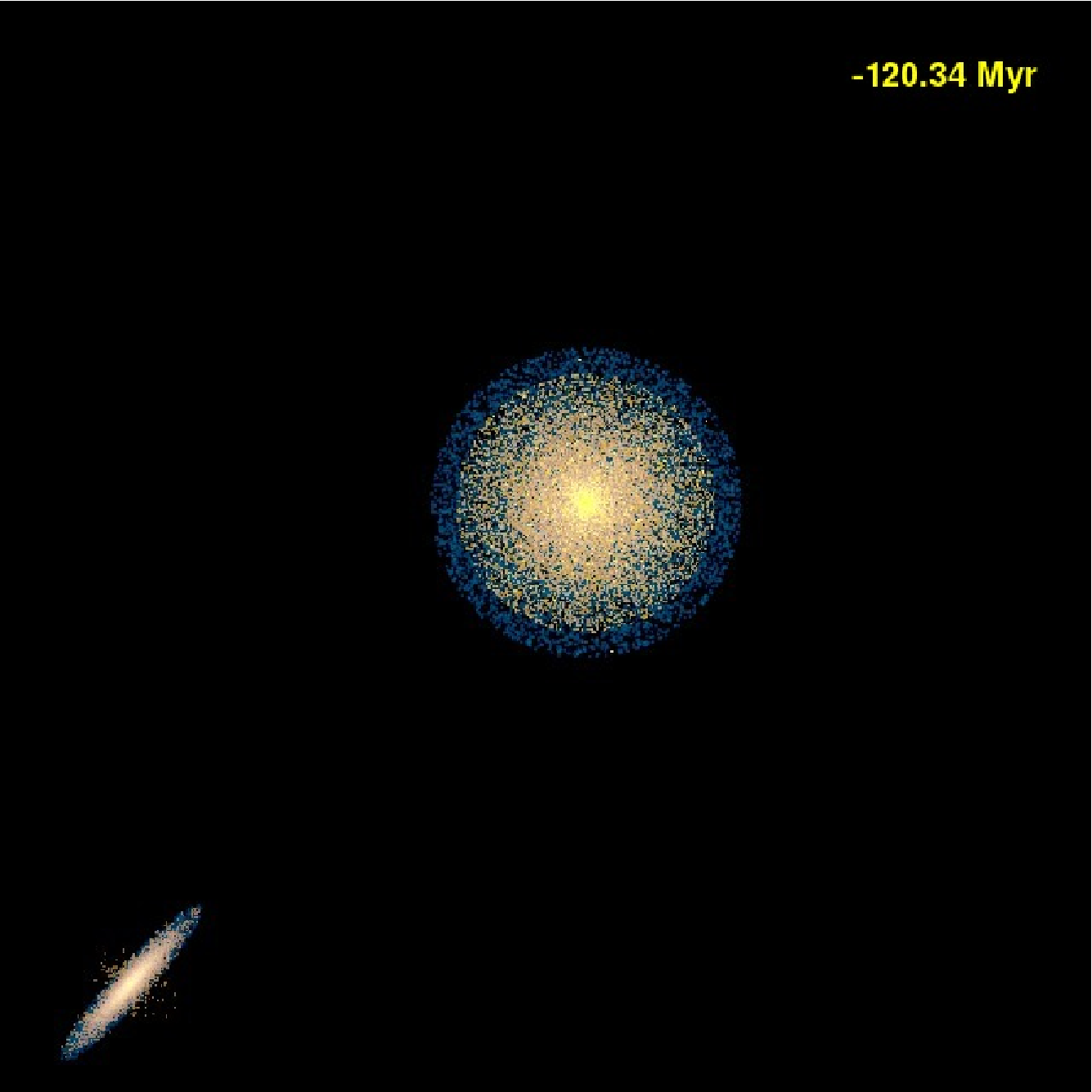}}
	\fbox{\includegraphics[width=\columnwidth]{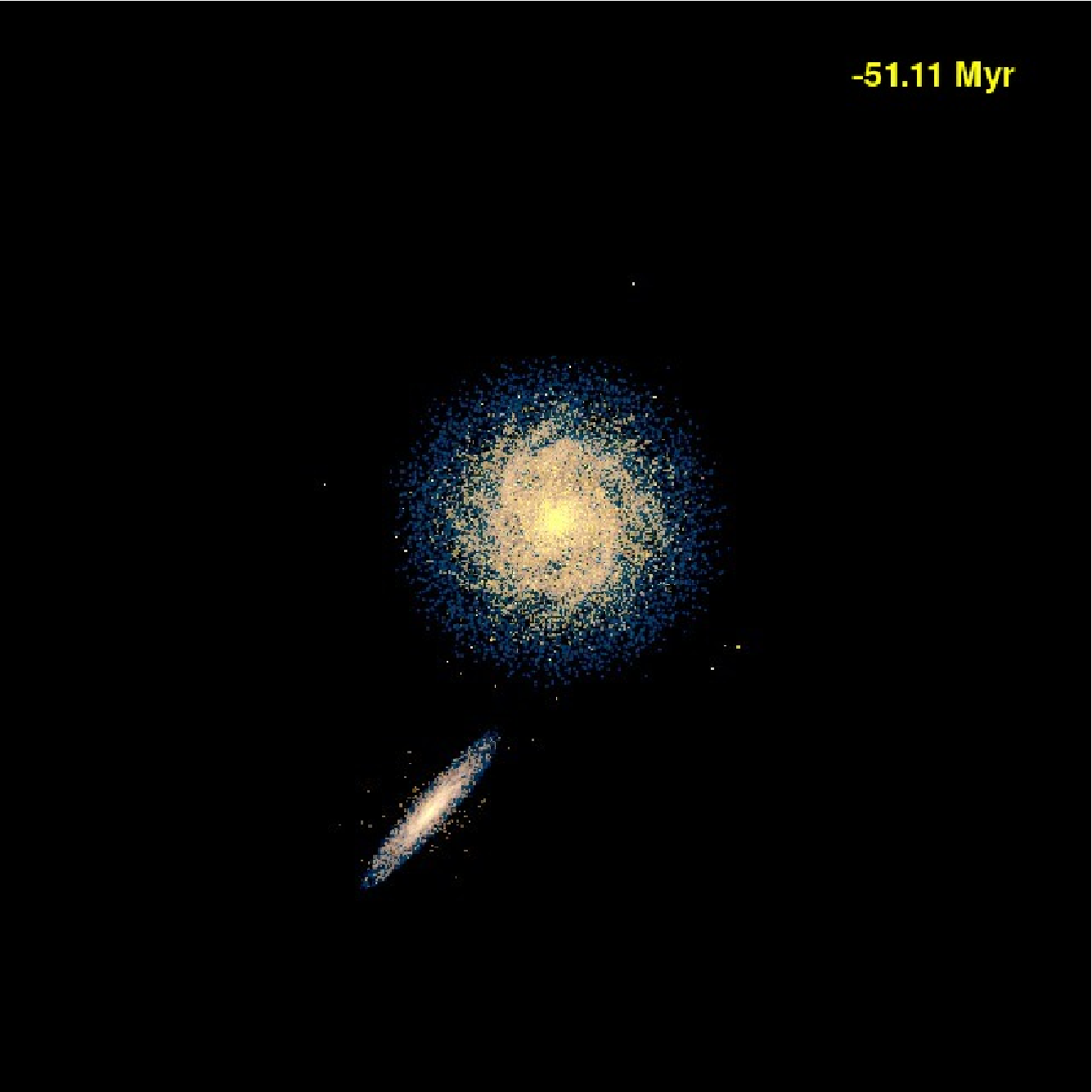}}
	\fbox{\includegraphics[width=\columnwidth]{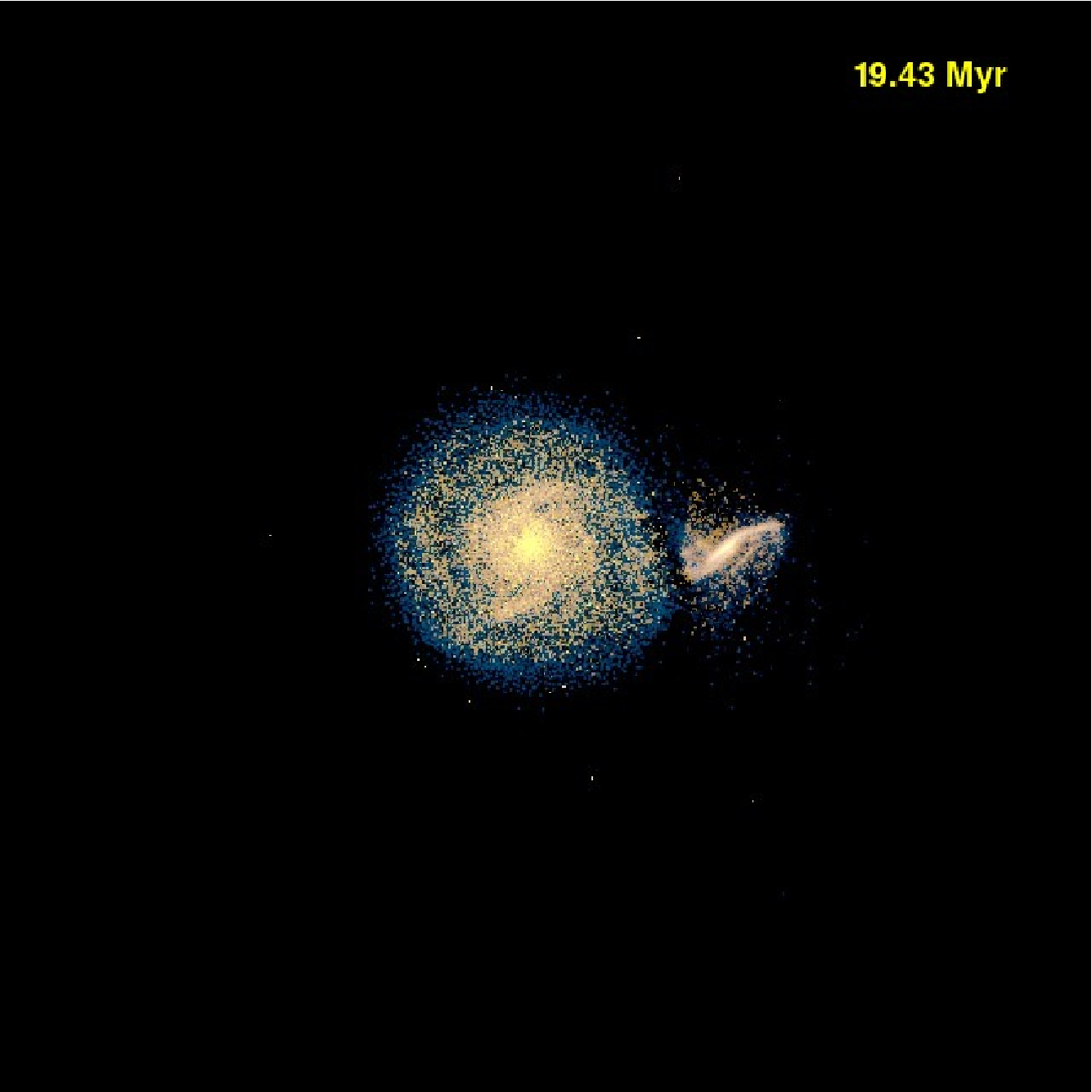}}
	\fbox{\includegraphics[width=\columnwidth]{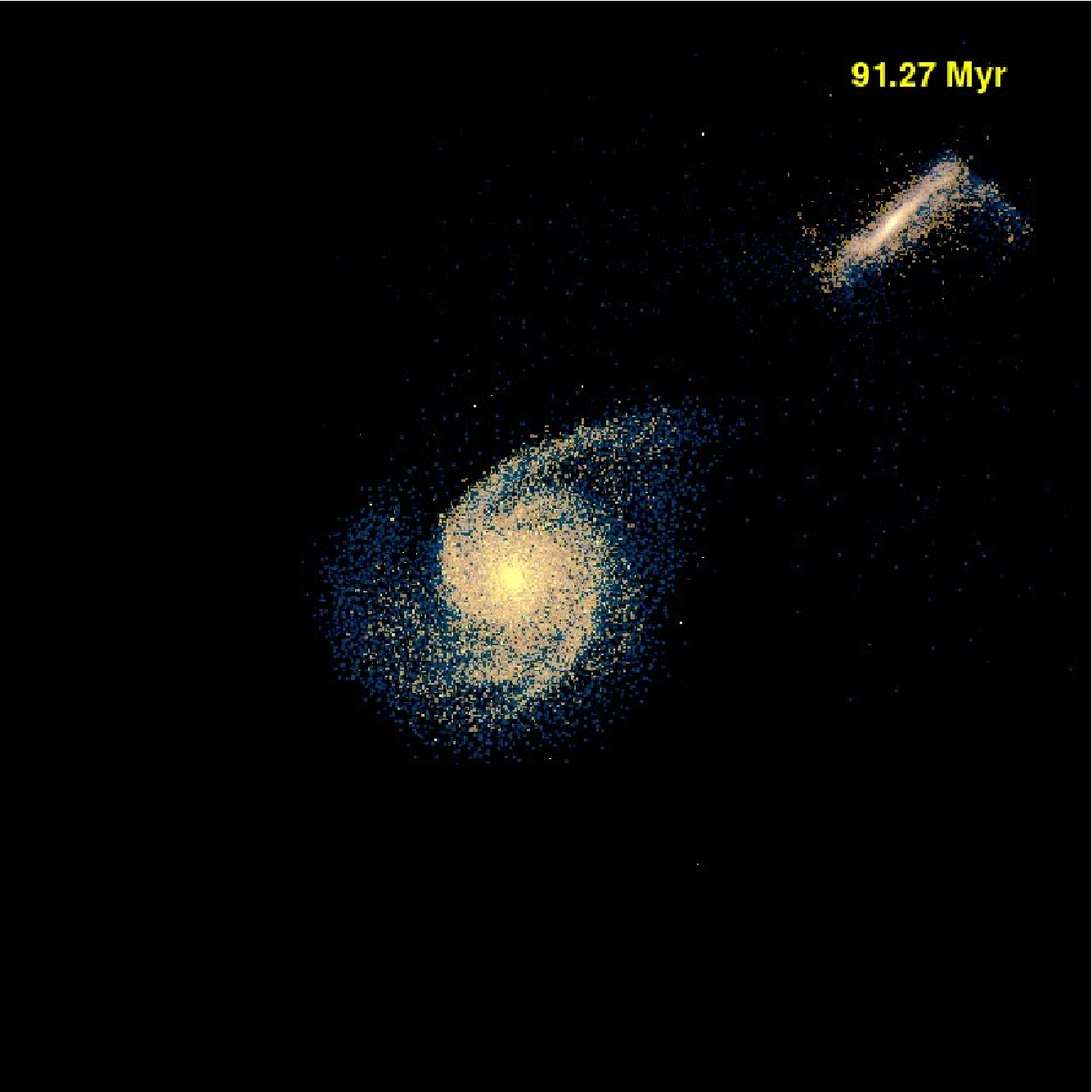}}
\caption{Time evolution of the encounter. Time is shown in Myr in the
upper right corner of each frame, with respect to the orbital
pericenter. The simulation begins at the upper left frame. Stars are
shown in light colors, while the gas is shown in blue. Dark matter
halo is not shown. Best fit to the current state of the AM\,2322-821
system is shown in the last frame.
}
	\label{simul}
\end{figure*}

Figure~\ref{simul} shows the time evolution of the encounter.  Time is shown in
Myr in the upper right corner of each frame, with respect to the
perigalacticum. Simulation starts $\sim$120\,Myr before perigalacticum. The
situation that best reproduces the morphology and kinematics of the present
stage of the AM\,2322-821 system is at $t=91.27$\,Myr after perigalacticum. The
general large scale morphology and kinematics agrees well with observations,
within the resolution provided by the simulations.\\ \\

\section{Stellar Population Synthesis}
\label{sintese}
A detailed study of the star formation in minor merger galaxies is an
important source of information not only on the age distribution of
their stellar population components, but to better understand several
aspects related to the interacting process, its effect in the
properties of the individual galaxies and their later evolution. The
absorption features arising from the  stellar component also affect to
different degrees the measured intensity of the emission line in the
spectrum of the gaseous component. This effect is more prominent in,
but not restricted to, the Balmer lines, so the stellar population
contribution must be subtracted from each  spectra in order to
study the physical properties of the gas.

\begin{figure*}
\centering
\includegraphics*[angle=-90,width=\columnwidth]{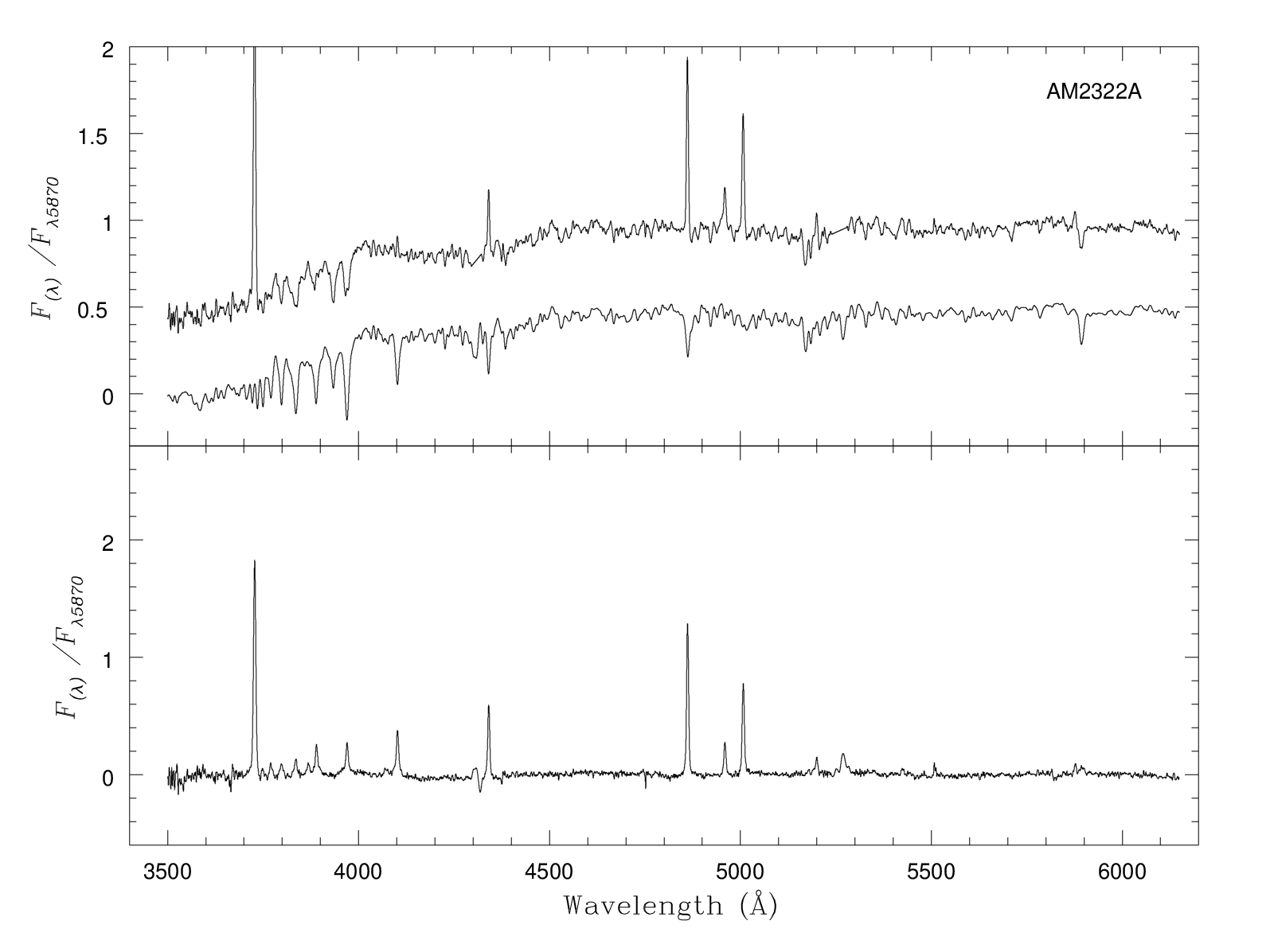}
\centering
\includegraphics*[angle=-90,width=\columnwidth]{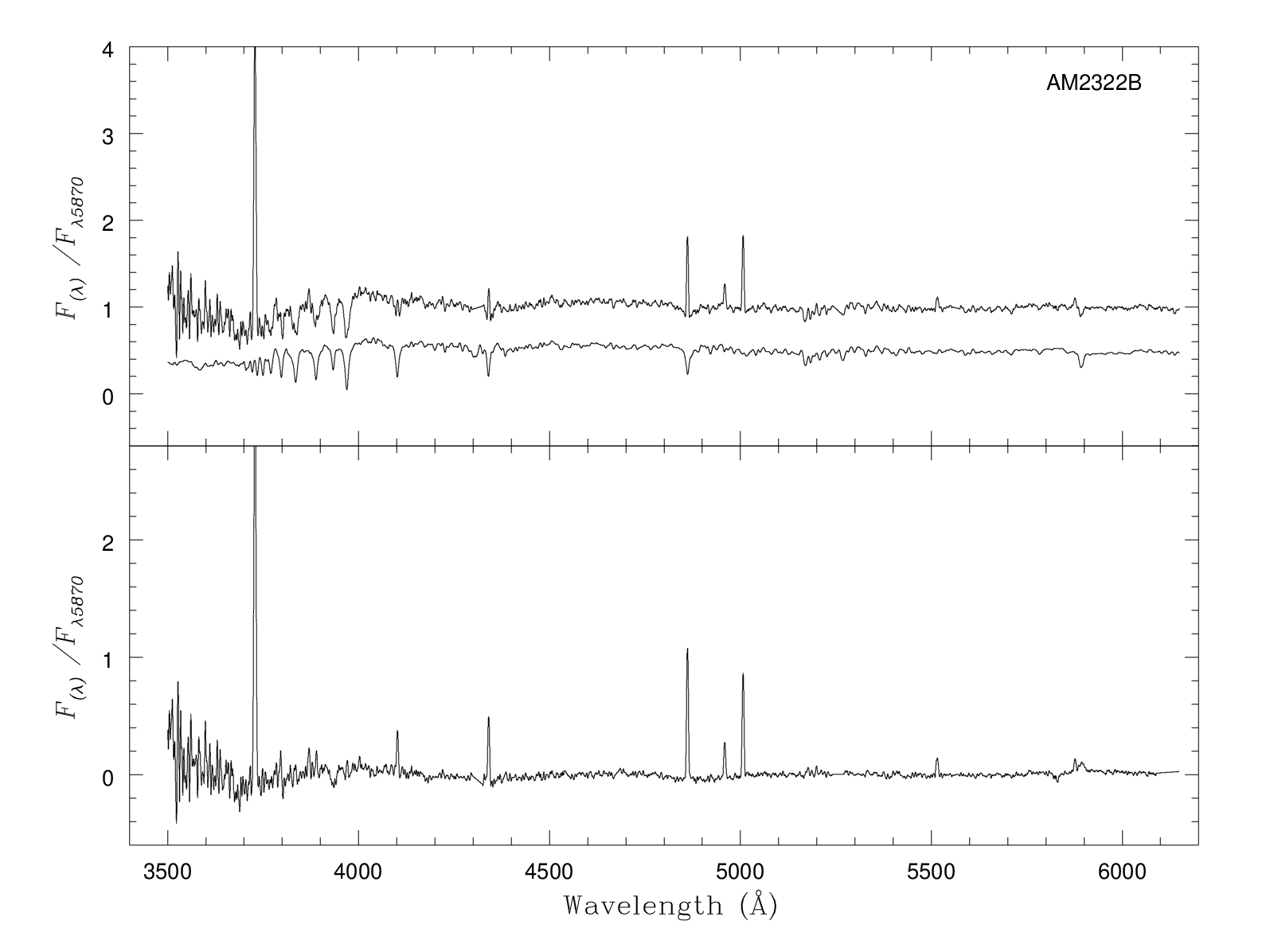}
\caption{Stellar population synthesis  for the central bin along the
PA=59\degr\, and PA=318\degr\ slit positions, 
for AM\,2322A (left)  and AM\,2322B (right), respectively. Top panel: spectrum
corrected for reddening  and the synthesized
spectrum (shifted by a constant). Bottom panel: pure emission spectrum.}
\label{sintesec}
\end{figure*}

\begin{figure*}
\subfigure{
\includegraphics*[angle=-90,width=\columnwidth]{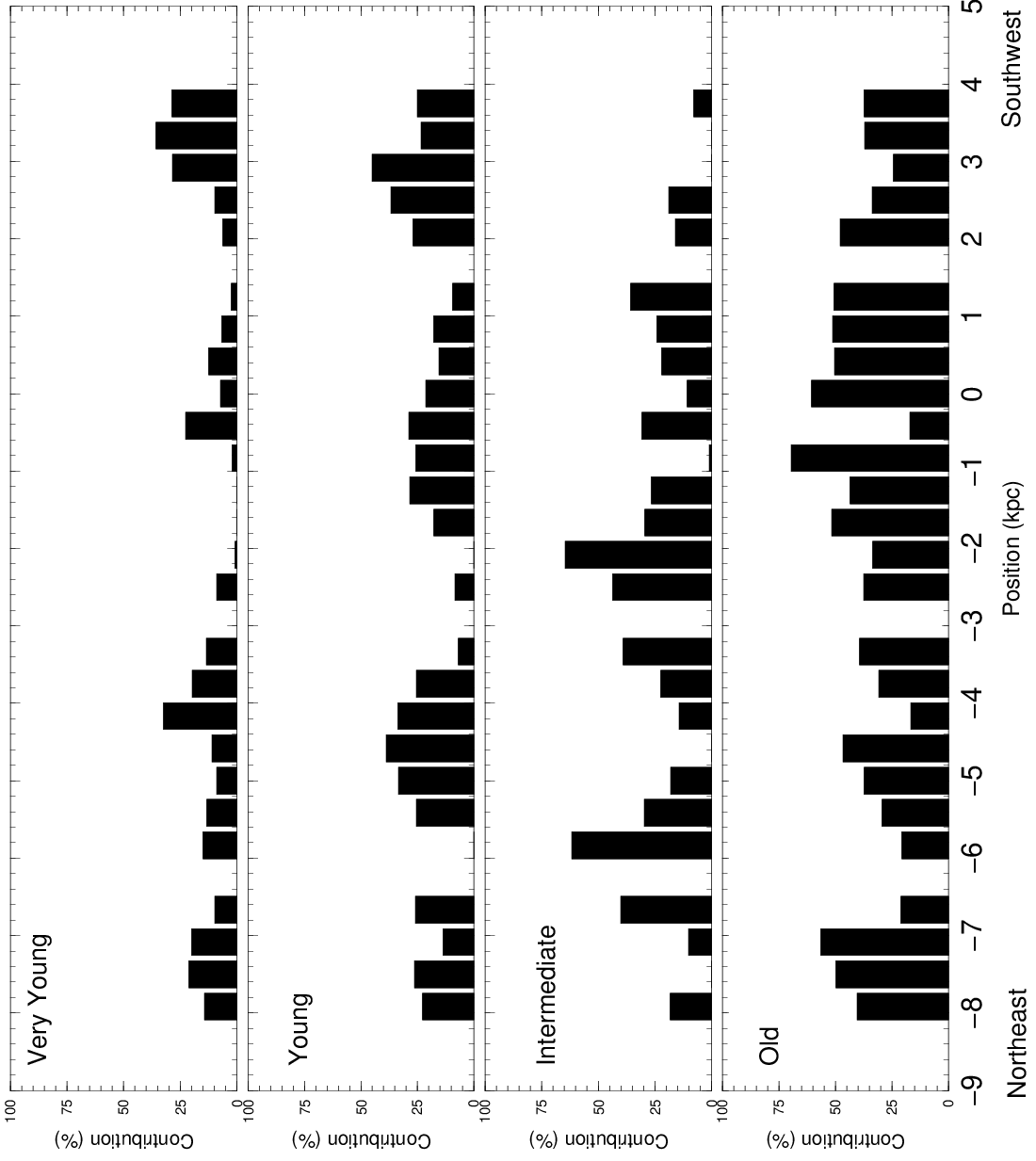}}
\subfigure{
\includegraphics*[angle=-90,width=\columnwidth]{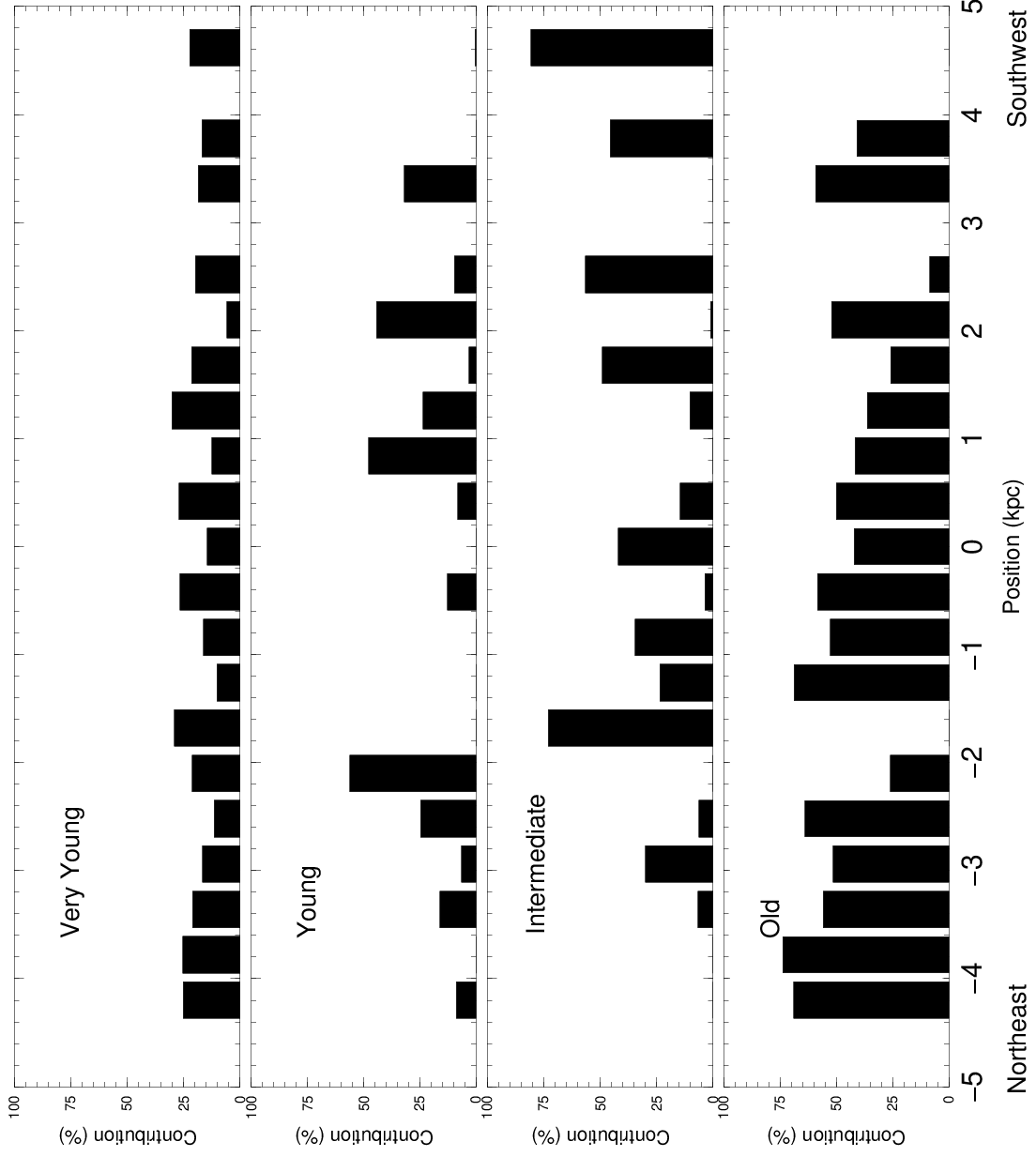}}
\caption{Synthesis results in flux fractions as a function 
of the distance to
the centre of the slit
along PA=59\degr\, (left) and PA=28\degr\, (right) for AM\,2322A.}
\label{synt_profile2}
\end{figure*}

To investigate the star formation history of AM\,2322A and AM\,2322B,
we use the stellar population synthesis code {\sc STARLIGHT} 
\citep{cid04,cid05,mateus06,asari07}.
This code is extensively discussed in \citet{cid04,cid05}, and
is built upon computational techniques originally developed for empirical
population synthesis with additional ingredients from evolutionary synthesis
models.The code fits an observed spectrum $O_{\lambda}$ with a combination of $N_{\star}$
single stellar populations (SSPs) from the \citet{bruzual03} models.
These models are based on a high-resolution library of observed stellar spectra,
which allows for detailed spectral evolution of the SSPs   at a resolution of 3\,\AA\, across the wavelength
range of 3\,200-9\,500 $\AA$
with a wide range of metallicities. We used the Padova 1994 tracks as
recommended by \citet{bruzual03}, with the 
initial mass function of Chabrier \citep{chabrier03} between 0.1 and 100
$M_{\sun}$.
Extinction is modeled by {\sc STARLIGHT} as due to foreground dust, using the
reddening law of \citet{cardelli89} with  R$_V$= 3.1 , and parametrized by the
V-band extinction  A$_V$.   The SSPs used in this work cover fifteen ages, 
t = [0.001\,, 0.003\,, 0.005\,, 0.01\,,
0.025\,, 0.04\,, 0.1\,,  0.3\,, 0.6\,, 0.9\,, 1.4\,, 2.5\,, 5\,, 11\,, and 13]
Gyr, and three metallicities, Z = [0.2 Z$_{\sun}$, 1 Z$_{\sun}$, and 2.5] Z$_{\sun}$, summing up  to 45 SSP components. Briefly, the
code solves the following equation for a model spectrum M$_{\sun}$ \citep{cid05}:

\begin{equation}
M_{\lambda} = M_{\lambda\,0}\biggl[\sum^{N_{\star}}_{j=1} x_{j} b_{j},_{\lambda} r_{\lambda}\biggr] G(v_{\star}, 
\sigma_{\star}) 
\end{equation}
were $b_{j},_{\lambda} r_{\lambda}$ is the reddened spectrum of the ${j}$th SSP normalized at
$\lambda _{0}$; $r_{\lambda} = 10^{0.4(A_{\lambda}- A_{\lambda_0})}$ is the reddening term; M$_{\lambda 0}$ is the synthetic
flux at the normalisation wavelength; $\vec{x}$ is the population vector;
$\otimes$ denotes the convolution operator; and G($v_{\star}, \sigma_{\star}$) is the gaussian
distribution used to model the line-of-sight stellar motions, it
is centred at velocity $v_{\star}$ with dispersion $\sigma_{\star}$.

The fit is carried out with a simulated annealing plus Metropolis
scheme, which searches for the minimum of the equation \citep{cid05}:

\begin{equation}
\chi^{2} = \sum_{\lambda}[(O_{\lambda} - M_{\lambda)}w_{\lambda}]^{2} 
\end{equation}
where emission lines and spurious features are masked out by fixing
$w_{\lambda}=0$. For more details on STARLIGHT see \citet{cid05}.

\begin{figure}
\includegraphics*[angle=-90,width=\columnwidth]{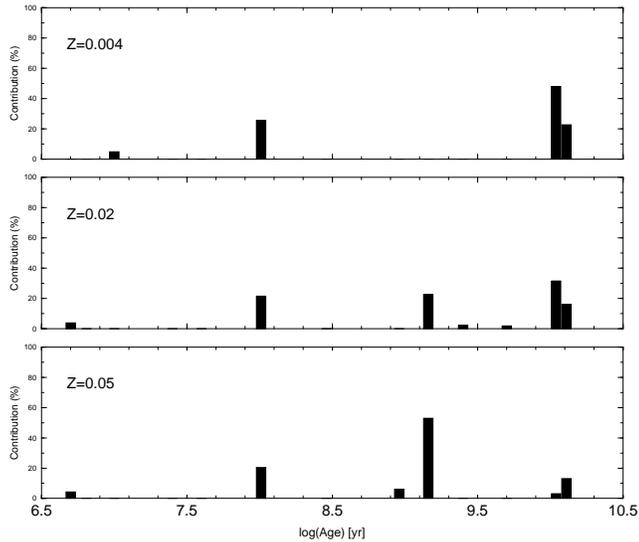}
\caption{Histogram for the central region along PA=59\degr\, showing the contribution to the flux continuum
(in percentage of each stellar age component for the 3 metallicity values considered).}
\label{grafico4}
\end{figure}

Prior to the modelling, the SSPs models were convolved to the same resolution of the observed spectra;
the observed spectra were shifted to its rest-frame, corrected for foreground Galactic reddening of $E(B-V)=0.181$ mag taken from
\citet{schlegel98} and normalized to $\lambda\,5870$\AA. The error in $O_{\lambda}$ considered in the fitting was the 
continuum rms with a $S/N  \ge 10$, where $S/N$ is the signal-to-noise  ratio per $\AA$ in the region around $\lambda_{0}=5870\,
\AA$. Also, the fitting was performed only in spectra with the presence of absorption lines.
 Measurement errors are still a problem in population synthesis.  
The most serious drawback of {\sc STARLIGHT}, as it is, is that it does not provide error
estimates on its parameters.  The reliability of parameter
estimation was best studied in \citet{cid04,cid05} by means of simulations
which fed the code with spectra generated with known parameters,
add noise, and then examined the correspondence
between input and output values. They performed this kind of simulation for a $N_{\star}$= 20
base and the main results of that study were:
(1) in the absence of noise, the method recovers all components
of $\vec{x}$ to a high degree of accuracy; (2) in the presence
of noise, however, the individual output $x_{j}$ fractions may
deviate drastically from the input values. 
However, this problem can be circumvented  by binning the stellar populations according to the
flux contributions. In addition to, the  absorption features present in our spectra, i.e. Balmer lines, 
are mainly dependent on age rather than on metallicity. 
In Fig. \ref{grafico4} is a histogram with the contribution of each stellar populations 
to the optical flux $\lambda\, 5\,870\, \AA\,$ of the spectrum of the central region along the PA=59\degr\,,  
before the binning the stellar populations. About the same results were obtained for the
other apertures. It can be noted in Fig. \ref{grafico4} that the about the same contributions by age are obtained from different metallicities, with a little increase of old population contribution when lower Z values are assumed (the metallicity-age degeneracy problem).

\begin{figure}
\includegraphics*[angle=-90,width=\columnwidth]{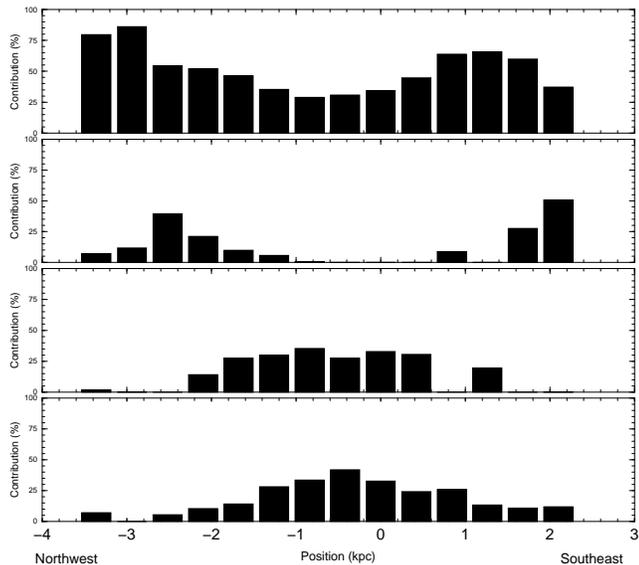}
\caption{Same as \ref{synt_profile2}, but  
along PA=318\degr\, of AM\,2322B}
\label{synt_profile1}
\end{figure}

Fig. \ref{sintesec} shows an example of the observed
spectra corrected by reddening, the model stellar population spectra and the
pure emission spectra for AM\,2322A and AM\,2322B.
The results of the synthesis are summarized in Tables \ref{synt_b} and 
\ref{synt_c} for the individual spatial bins in each galaxy,
stated as the percentual contribution of each base element to the flux  at
$\lambda\, 5\,870\, \AA$.

Following the prescription of \citet{cid05}, we defined a condensed population
vector, 
by binning the stellar populations according to the flux contributions 
into very young, $x_{\rm VY}$ (t $ \leq 1\times10^{8}$ yr), 
young, $x_{\rm Y}$ ($ 1\times10^{8} <\rm t \leq 5\times10^{8}$ yr),
intermediate-age,  $x_{\rm I}$ ($ 5\times10^{8} <\rm t \leq 2\times10^{9}$ yr)
and 
old, $x_{\rm O}$ (t $ > 2\times10^{9}$ yr) components. The same bins were used
to represent the mass components of the population vector ($m_{\rm VY}$,
$m_{\rm Y}$, $m_{\rm I}$ and
$m_{\rm O}$).   The metallicity, an important parameter to characterize the stellar population content,
is weighted by light fraction, and the  results point out to a mean value around
solar.  The quality of the fitting result is measured by the parameters 
$\chi^{2}$ and  $adev$. The latter gives the percentual mean deviation
$|O_{\lambda} - M_{\lambda}|/O_{\lambda}$ over all fitted
pixels, where $O_{\lambda}$ and $M_{\lambda}$ are the observed and model
spectra, respectively.

The spatial variation in the contribution of the stellar-population components
are shown in \ref{synt_profile2} and \ref{synt_profile1} for the galaxies 
AM\,2322A and AM\,2322B, respectively.

As can be seen in  Fig. \ref{synt_profile2}, the stellar population of AM\,2322A
is heterogenous along  both slit positions. It does not show age gradient: the
very young $x_{\rm VY}$, intermediate  $x_{\rm I}$ and old populations $x_{\rm
O}$ contribute significantly to the optical flux at 
$\lambda\, 5\,870\, \AA$.  In a mass-weighted context, the bulge is  predominantly
composed of old population, in agreement with the results obtained by \citep{macarthur09}.

AM\,2322B is dominated, by the very young population component $x_{\rm VY}$.
This component has   
a systematic variation along the slit, increasing outwards. Instead  the
fraction of old and intermediate stellar population
components   is decreasing from the centre to the outer regions.
The young star-formation episode ocurred  in this galaxy
could be related with the perigalactic passage, that is, about 90 Myr after
perigalacticum.
The effects of the galaxy interaction in the star formation history are much 
more conspicuous
in the secondary galaxy than in the main galaxy. This is expected as the
triggering of star formation events by tidal interactions appears to be more
efficient in less massive systems. Similar results were  also found  for the
pair AM\,2306-721 \citep{krabbe08}.


\section{Gas phase O/H abundance}
\label{emission}
Once the stellar population contribution has been determined, the
underlying absorption line spectrum can be subtracted to allow the
measurement and analysis of the line emission from the gaseous
component. The line intensities were measured using Gaussian line
profile fitting on the pure emission spectra. We used the {\sc IRAF splot}
routine to fit
the lines, with the associated error being given as $\sigma^{2} =
\sigma_{cont}^{2} + \sigma_{line}^{2} $, where $\sigma_{cont}$ and
$\sigma_{line}$ are the continuum 
rms and the Poisson error of the line flux, respectively. 
  The relative error in the flux estimates of the emission lines´
are lower than 20 \%.
The residual
extinction associated with the gaseous component for each spatial bin
was calculated comparing the observed $\rm H\gamma/H\beta$ and $\rm H\alpha/H
\beta$ ratios to the 
theoretical values in \citet{hummer87} for an electron temperature of 10\,000
$\mathrm{K}$ and a electron density of
100 $\mathrm{cm^{-3}}$. The observed emission line intensities were then
corrected by this residual extinction using the
\citet{howarth83} reddening function. 

\begin{table*}
\caption{Stellar-population synthesis results for AM\,2322A}
\label{synt_b}
\begin{tabular}{lrrrrrrrrrrrrr}
 \\
\noalign{\smallskip}
\hline
\hline
\noalign{\smallskip}
Pos. (kpc) & \multicolumn{1}{c}{$x_{\rm VY}$} & \multicolumn{1}{c}{$ x_{\rm Y}$}
&
 \multicolumn{1}{c}{$ x_{\rm I}$} &  \multicolumn{1}{c}{$x_{\rm O}$}
 & \multicolumn{1}{c}{$ m_{\rm VY}$} & 
 \multicolumn{1}{c}{$ m_{\rm Y}$}
  & \multicolumn{1}{c}{$ m_{\rm I}$} &
 \multicolumn{1}{c}{$m_{\rm O}$}& 
  \multicolumn{1}{c}{$Z_{\star}$[1]} &
 \multicolumn{1}{c}{$ \chi^{2}$} & 
 \multicolumn{1}{c}{$\rm adev$} & \multicolumn{1}{c}{$\rm A_{v}$}
 \\
& \multicolumn{1}{c}{(\%)} & \multicolumn{1}{c}{(\%)} &
 \multicolumn{1}{c}{(\%)} &  \multicolumn{1}{c}{(\%)}
 & \multicolumn{1}{c}{(\%)} & 
 \multicolumn{1}{c}{(\%)}
  & \multicolumn{1}{c}{(\%)} &
 \multicolumn{1}{c}{(\%)}& 
  \multicolumn{1}{c}{} &
 \multicolumn{1}{c}{} & 
 \multicolumn{1}{c}{} & \multicolumn{1}{c}{(mag)}
 \\
\hline
\noalign{\smallskip}
\multicolumn{13}{c}{AM\,2322A (PA=59\degr)}\\
\noalign{\smallskip}
\hline
\noalign{\smallskip}
 -7.98   &   14.3  &   22.8  &   18.2  &   40.6&   0.4  &   2.3  &   3.1  &  94.2  &	0.014  &    1.4  &  5.83  &  0.15 \\
 -7.56   &   21.1  &   26.3  &   0.0   &   49.9&   0.5  &   1.8  &   0.0  &  97.7  &	0.028  &    0.1  &  7.98  &  0.00 \\
 -7.14   &   19.9  &   13.7  &   10.2  &   56.6&   0.8  &   1.8  &   5.7  &  91.6  &	0.010  &    0.1  &  24.79 &  0.00 \\
 -6.72   &   9.7   &   25.9  &   40.2  &   21.2&   0.2  &   4.0  &   12.4 &  83.4  &	0.024  &    0.1  &  10.96 &  0.00 \\
 -5.88   &   14.8  &   0.0   &   61.8  &   20.7&   0.3  &   0.0  &   37.0 &  62.7  &	0.007  &    0.1  &  13.14 &  0.42 \\
 -5.46   &   13.2  &   25.6  &   29.6  &   29.4&   0.1  &   3.8  &   10.2 &  85.9  &	0.019  &    0.0  &  10.20 &  0.48 \\
 -5.04   &   8.8   &   33.4  &   18.1  &   37.4&   0.1  &   3.3  &   3.5  &  93.1  &	0.016  &    0.1  &  7.65  &  0.24 \\
 -4.62   &   10.9  &   38.9  &   0.0   &   46.8&   0.1  &   2.2  &   0.0  &  97.7  &	0.021  &    0.1  &  5.59  &  0.25 \\
 -4.20   &   32.3  &   33.7  &   14.2  &   16.8&   1.7  &   10.7 &   9.4  &  78.2  &	0.022  &    0.1  &  4.88  &  0.34 \\
 -3.78   &   19.7  &   25.5  &   22.4  &   31.0&   0.4  &   2.0  &   5.8  &  91.8  &	0.033  &    0.1  &  4.95  &  0.12 \\
 -3.36   &   13.4  &   6.9   &   39.1  &   39.5&   0.1  &   0.7  &   7.9  &  91.2  &	0.022  &    0.1  &  5.54  &  0.18 \\
 -2.52   &   8.8   &   8.5   &   43.7  &   37.6&   0.1  &   0.9  &   10.3 &  88.8  &	0.021  &    1.2  &  3.88  &  0.38 \\
 -2.10   &   0.7   &   0.0   &   64.6  &   33.7&   0.0  &   0.0  &   11.9 &  88.1  &	0.021  &    0.1  &  12.37 &  0.05 \\
 -1.68   &   0.0   &   17.8  &   29.5  &   51.8&   0.0  &   0.8  &   4.6  &  94.5  &	0.017  &    0.2  &  22.41 &  0.21 \\
 -1.26   &   0.0   &   28.4  &   26.5  &   43.7&   0.0  &   1.5  &   5.1  &  93.4  &	0.031  &    0.1  &  19.28 &  0.51 \\
-0.84    &   1.9   &   25.8  &   1.0   &   69.7&   0.0  &   1.3  &   0.2  &  98.5  &	0.025  &    0.0  &  15.76 &  0.78 \\
-0.42    &   22.6  &   28.8  &   30.8  &   17.2&   2.1  &   5.4  &   21.0 &  71.6  &	0.018  &    0.1  &  7.19  &  1.40 \\
    0    &   7.1   &   21.3  &   10.7  &   60.7&   0.1  &   1.1  &   2.9  &  95.9  &	0.023  &    0.0  &  2.77  &  0.46 \\
 0.42    &   12.5  &   15.5  &   21.9  &   50.6&   0.1  &   0.8  &   7.8  &  91.3  &	0.024  &    0.0  &  3.25  &  0.39 \\
 0.84    &   6.5   &   17.9  &   24.0  &   51.4&   0.0  &   0.9  &   7.3  &  91.7  &	0.028  &    0.1  &  8.98  &  0.53 \\
  1.26   &   2.3   &   9.4   &   35.7  &   50.8&   0.1  &   0.7  &   9.2  &  90.1  &	0.029  &    0.1  &  16.82 &  0.46 \\
  2.10   &   6.2   &   27.0  &   16.0  &   48.1&   0.1  &   2.4  &   2.7  &  94.8  &	0.021  &    0.2  &  31.91 &  0.24 \\
  2.52   &   9.7   &   36.7  &   18.8  &   33.9&   0.1  &   2.3  &   3.3  &  94.3  &	0.020  &    0.2  &  12.53 &  0.29 \\
  2.94   &   28.4  &   45.1  &   0.0   &   24.4&   0.5  &   3.6  &   0.0  &  95.9  &	0.015  &    0.2  &  7.64  &  0.35 \\
  3.36   &   35.8  &   23.5  &   0.0   &   37.2&   0.3  &   1.6  &   0.0  &  98.1  &	0.032  &    0.2  &  7.87  &  0.23 \\
  3.78   &   28.7  &   25.0  &   7.8   &   37.3&   0.6  &   4.3  &   2.2  &  92.9  &	0.029  &    0.2  &  19.06 &  0.22 \\
\noalign{\smallskip}

\hline
\noalign{\smallskip}
\multicolumn{13}{c}{AM\,2322A (PA=28\degr)}\\
\noalign{\smallskip}
\hline
\noalign{\smallskip}
 -4.20        &   25.2  &   8.9  &   0.0   &   69.0&   0.9  &   1.2  &	0.0    &  97.8  &   0.012  &    0.1  &  16.85  &  0.17 \\
 -3.78        &   25.3  &   0.0  &   0.0   &   73.8&   0.4  &   0.0  &	0.0    &  99.6  &   0.004  &    0.1  &  21.13  &  0.00 \\
 -3.36        &   21.0  &   16.1 &   6.6   &   56.0&   0.1  &   1.5  &	1.1    &  97.2  &   0.013  &    0.1  &  11.01  &  0.09 \\
 -2.94        &   16.6  &   6.6  &   29.9  &   51.6&   0.1  &	0.6  &   11.5  &  87.8  &   0.012  &    0.1  &   9.93  &  0.10 \\
 -2.52        &   11.4  &   24.6 &   6.1   &   64.1&   0.3  &   2.7  &	3.3    &  93.7  &   0.010  &    0.1  &   9.77  &  0.28 \\
 -2.1         &   21.2  &   56.2 &   0.0   &   26.1&   0.4  &   7.4  &	0.0    &  92.2  &   0.026  &    0.1  &   9.64  &  0.53 \\
 -1.68        &   29.0  &   0.0  &   73.0  &   0.0 &   2.5  &   0.0  &	97.3   &  0.2   &   0.022  &    0.1  &   7.99  &  0.41 \\
 -1.26        &   10.1  &   0.0  &   23.5  &   68.8&   0.2  &	0.0  &   8.2   &  91.5  &   0.009  &    0.1  &  11.85  &  0.19 \\
 -0.84        &   16.2  &   0.0  &   34.5  &   52.9&   0.5  &	0.0  &   8.6   &  90.9  &   0.007  &    0.1  &   8.10  &  0.37 \\
 -0.42        &   26.6  &   12.8 &   3.5   &   58.4&   0.4  &   1.4  &	0.9    &  97.3  &   0.020  &    0.1  &   7.93  &  0.12 \\
   0          &   14.4  &   0.0  &   41.9  &   42.2&   0.6  &	0.0  &   14.3  &  85.1  &   0.012  &    0.1  &   8.07  &  0.42 \\
  0.42        &   26.9  &   8.2  &   14.4  &   50.0&   0.3  &	0.8  &   4.3   &  94.6  &   0.018  &    0.1  &   7.50  &  0.45 \\
  0.84        &   12.5  &   47.9 &   0.0   &   41.8&   0.4  &   3.6  &	0.0    &  96.0  &   0.026  &    0.1  &   6.90  &  0.54 \\
  1.26        &   30.2  &   23.6 &   10.1  &   36.3&   0.5  &	4.2  &   5.8   &  89.5  &   0.015  &    0.1  &   7.32  &  0.68 \\
  1.68        &   21.3  &   3.3  &   49.0  &   26.0&   0.3  &	0.6  &   37.0  &  62.2  &   0.016  &    0.1  &   9.91  &  0.56 \\
  2.10        &   5.8	&   44.3 &   0.9   &   52.2&   0.2  &   5.4  &	0.3    &  94.1  &   0.007  &    0.1  &  39.77  &  0.76 \\
  2.52        &   19.6  &   9.7  &   56.6  &   8.6 &   0.8  &   2.2  &	46.1   &  51.0  &   0.016  &    0.1  &  18.30  &  0.80 \\
  3.36        &   18.4  &   31.9 &   0.0   &   59.3&   0.5  &   5.9  &	0.0    &  93.6  &   0.019  &    0.1  &  22.39  &  0.00 \\
  3.78        &   16.8  &   0.0  &   45.6  &   41.0&   0.7  &	0.0  &   21.4  &  77.9  &   0.022  &    0.1  &  30.54  &  0.23 \\
  4.62        &   22.2  &   0.6  &   80.7  &   0.1 &   1.6  &   0.2  &	97.9   &  0.2   &   0.038  &    0.1  &  54.97  &  0.66 \\
\noalign{\smallskip}
\hline
\noalign{\smallskip}
\noalign{\smallskip}
\noalign{\smallskip}
\end{tabular}
\begin{minipage}[c]{18.0cm}
[1] Abundance by mass with Z$_{\sun}$=0.02 \\
\end{minipage}
\end{table*}

\begin{table*}
\caption{Stellar-population synthesis results for AM\,2322B}
\label{synt_c}
\begin{tabular}{lrrrrrrrrrrrr}
 \\
\noalign{\smallskip}
\hline
\hline
\noalign{\smallskip}

Pos. (kpc) & \multicolumn{1}{c}{$x_{\rm VY}$} & \multicolumn{1}{c}{$ x_{\rm Y}$}
&
 \multicolumn{1}{c}{$ x_{\rm I}$} &  \multicolumn{1}{c}{$x_{\rm O}$}
 & \multicolumn{1}{c}{$ m_{\rm VY}$} & 
 \multicolumn{1}{c}{$ m_{\rm Y}$}
  & \multicolumn{1}{c}{$ m_{\rm I}$} &
 \multicolumn{1}{c}{$m_{\rm O}$}& 
  \multicolumn{1}{c}{$Z_{\star}$[1]} &
 \multicolumn{1}{c}{$ \chi^{2}$} & 
 \multicolumn{1}{c}{$\rm adev$} & \multicolumn{1}{c}{$\rm A_{v}$}
 \\
& \multicolumn{1}{c}{(\%)} & \multicolumn{1}{c}{(\%)} &
 \multicolumn{1}{c}{(\%)} &  \multicolumn{1}{c}{(\%)}
 & \multicolumn{1}{c}{(\%)} & 
 \multicolumn{1}{c}{(\%)}
  & \multicolumn{1}{c}{(\%)} &
 \multicolumn{1}{c}{(\%)}& 
  \multicolumn{1}{c}{} &
 \multicolumn{1}{c}{} & 
 \multicolumn{1}{c}{} & \multicolumn{1}{c}{(mag)}
 \\
\hline
\noalign{\smallskip}
\multicolumn{13}{c}{AM\,2322B (PA=318\degr)}\\
\noalign{\smallskip}
\hline
\noalign{\smallskip}
-3.36   &   79.7  &   7.1   &	1.8   &   7.1  &   5.5  &   2.3   &   1.3  &  91.0  &   0.021  &     1.30  &  3.94  &  0.37 \\
-2.94   &   86.1  &   11.8  &   0.2   &   0.0  &   47.3 &   49.4  &   3.2  &  0.0   &   0.020  &     0.01  &  4.01  &  0.00 \\
-2.52   &   54.4  &   39.4  &   0.0   &   5.6  &   4.0  &   13.9  &   0.0  &  82.1  &   0.022  &     0.01  &  3.49  &  0.16 \\
-2.10   &   51.9  &   21.2  &   14.1  &   10.6 &   2.3  &   5.3   &   7.7  &  84.6  &	0.026  &     0.01  &  3.49  &  0.10 \\
-1.88   &   46.3  &   9.9   &	27.5  &   14.2 &   2.6  &   2.9   &   15.6 &  78.9  &	0.022  &     0.01  &  3.26  &  0.00  \\
-1.26   &   35.4  &   5.7   &	29.9  &   28.3 &   0.5  &   0.9   &   7.1  &  91.6  &   0.018  &     0.01  &  2.77  &  0.00 \\
-0.84   &   28.8  &   0.8   &	35.4  &   33.7 &   0.4  &   0.1   &   8.2  &  91.3  &	0.019  &     0.01  &  2.71  &  0.12 \\
-0.42   &   30.7  &   0.1   &	27.3  &   42.0 &   0.4  &   0.0   &   5.0  &  94.6  &   0.022  &     0.01  &  2.74  &  0.28 \\
0       &   34.2  &   0.0   &	32.9  &   32.9 &   0.5  &   0.0   &   6.9  &  92.7  &	0.021  &     0.01  &  2.67  &  0.54 \\
0.42    &   44.7  &   0.0   &	30.6  &   24.2 &   0.9  &   0.0   &   7.8  &  91.3  &	0.022  &     0.01  &  2.67  &  0.53 \\
0.84    &   63.8  &   8.8   &	0.0   &   26.0 &   1.7  &   2.2   &   0.0  &  96.2  &   0.025  &     0.01  &  2.83  &  0.46 \\
1.26    &   65.8  &   0.0   &	19.5  &   13.5 &   2.3  &   0.0   &   7.1  &  90.6  &	0.021  &     0.01  &  3.02  &  0.32 \\
1.68    &   59.9  &   27.6  &   0.0   &   10.9 &   5.3  &   11.2  &   0.0  &  83.5  &	0.018  &     0.01  &  2.96  &  0.32 \\
2.10    &   37.1  &   50.7  &   0.1   &   11.9 &   3.2  &   21.6  &   0.3  &  75.0  &	0.018  &     0.01  &  3.89  &  0.53 \\
\hline
\noalign{\smallskip}
\noalign{\smallskip}
\noalign{\smallskip}
\end{tabular}
\begin{minipage}[c]{18.0cm}
[1] Abundance by mass with Z$_{\sun}$=0.02 \\
\end{minipage}
\end{table*}

Determining accurate element abundances   of the gas phase from optical
spectra is
critically dependent on measuring temperature sensitive line ratios, such as
[\ion{O}{iii}]$(\lambda\,4959+\lambda\,5007)/\lambda\,4363$.
However, when doing spectroscopy of \ion{H}{ii} regions with high
metallicity and/or low excitation, temperature sensitive lines such as
[\ion{O}{iii}]$\lambda\,4363$ are found to be weak or unobservable,
and empirical indicators based on more easily measured  line ratios
have to be used to estimate metal abundances.

Because we did not detect any temperature-sensitive emission lines in our spectra,
we used the  $R_{23}$=([\ion{O}{ii}]$\lambda\,3727+$[\ion{O}{iii}]
$\lambda\,4959+$[\ion{O}{iii}] 
$\lambda\,5007)/$H$\beta$ vs.
[\ion{O}{iii}]$\lambda\,5007$/[\ion{O}{ii}]$\lambda\,3727$ \citep{mcgaugh91} 
diagnostic diagram  
to estimate the metallicity of the gas in the star forming regions by
comparing the observed values with  a grid of
photoionization models. The
[\ion{N}{ii}]$\lambda\,6584$/[\ion{o}{ii}]$\lambda\,3727$ line ratio was
used to break the $R_{23}$ degeneracy for our data, as suggested by
\citep{kewley08} .
We verify that  all observed  star forming regions
of our galaxies have
log([\ion{N}{ii}]$\lambda\,6584$/[\ion{o}{ii}]$\lambda\,3727$ $> -1$,
thus they are placed in the upper $R_{23}$ branch.

\begin{figure}
\centering
\includegraphics*[angle=-90,width=\columnwidth]{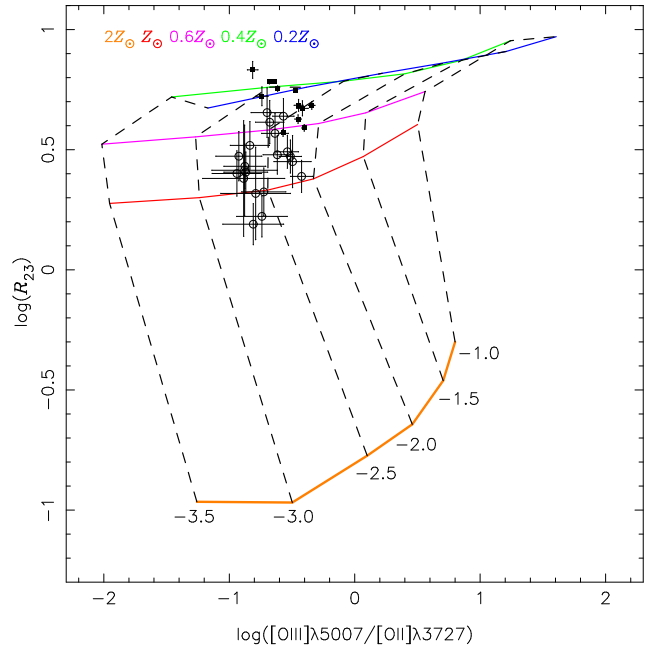}			
\caption{The relation $\log(R_{23})$ vs. $\log$([\ion{O}{iii}]/[\ion{O}{ii}])
for the individual spatial bins in
AM\,2322B (filled squares) and AM\,2322A (open triangles). The curves represent
the photoionization models described in the text (dashed lines
correspond to different values of the logarithm ionization parameter, solid
lines to different gas metallicities).}
\label{grade1}
\end{figure}

\begin{figure}
\centering
\includegraphics*[angle=-90,width=\columnwidth]{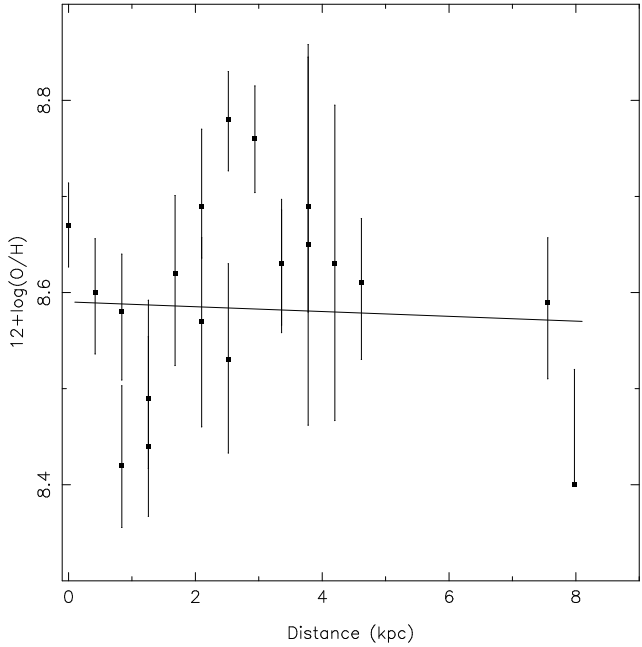}			
		     
\caption{The O/H abundance distribution as a function of the distance to the
center of the AM\,2322A
estimated using the $\log(R_{23})$ vs. $\log$([\ion{O}{iii}]/[\ion{O}{ii}]). The
solid line represents the linear fit of the data.}
\label{abund_a}
\end{figure}

\begin{figure}
\centering
\includegraphics*[angle=-90,width=\columnwidth]{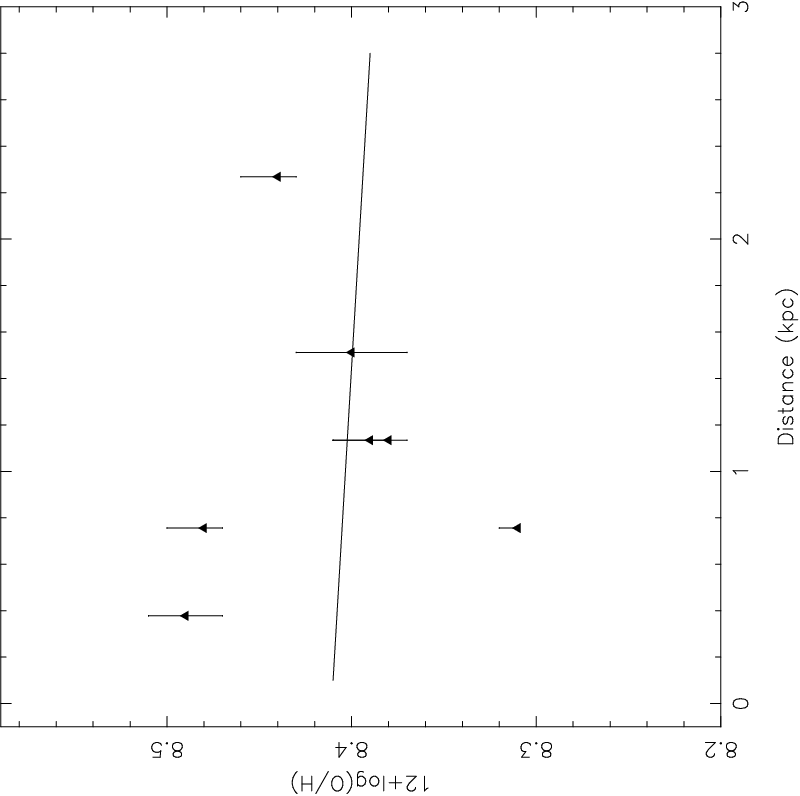}			
\caption{Same as Fig. \ref{abund_a}, but for AM\,2322B.}
\label{abund_b}
\end{figure}

The photoionization models were built
using the code {\sc Cloudy/08.00} \citep{ferland02}. In each model, a stellar
cluster
was assumed as the ionizing source with the stellar energy distribution
obtained from the synthesis models {\sc Starburst99} \citep{leitherer99}.
We  calculated models with  metallicities of $Z=$ 2.0,1.0, 0.6, 0.4, and 0.2
$Z_{\odot}$, 
ionization parameter $\log\,U=$ $-1.0$, $-1.5$, $-2$, $-2.5$, $-3.0$, and
$-3.5$, with the stellar cluster having an upper stellar  
mass limit of $M_{\rm up}$ = 100 $M_{\odot}$, age of 2.5 Myr and formed by an
instantaneous
burst. The solar value of $12+\ohlog =8.69$ 
is taken from \citet{allendre01}.  Similar models have been used to describe
observational data of \ion{H}{ii} regions in interacting
\citep{krabbe07,krabbe08} as well as in isolated
galaxies \citep{dors05}. The reader is referred to \citet{dors06} 
for a full description of the models. Fig. \ref{grade1} shows
the $R_{23}$ vs. [\ion{O}{iii}]$\lambda\,5007$/[\ion{O}{ii}]$\lambda\,3727$
diagram, with the observed values superposed in the grid of computed models.
Filled  squares correspond to regions in AM\,2322A and  open squares to regions
in AM\,2322B. 
Figs. \ref{abund_a} and \ref{abund_b} show the estimated O/H
abundance distribution as a function of the galactocentric distance  for
AM\,2322A and AM\,2322B, respectively.

As we can see in Figs. \ref{abund_a} and \ref{abund_b}, the oxygen abundances
derived from the $R_{23}$ vs. [O\,III]/[O\,II]
diagram suggest an relatively homogenous  O/H value of 8.6$(\pm 0.01)$ and 8.40$(\pm 0.05)$
across  the disc of the  galaxies AM\,2322A and AM\,2322B, respectively.


The central  O/H values of for the two galaxies are lower than those obtained
for field galaxies of the same luminosity, in agreement with the  results
obtained by \citet{kewley06} for galaxy pair members with small projected
separations (s $<$ 20 kpc).
This result provides strong observational
evidence for the hypothesis that the galaxy interactions create gas flows
towards the central regions,
carrying less enriched gas from the outskirts of the galaxy into the central
regions, mixing and 
homogenizing the chemical composition of the interstellar medium
\citep{kewley06}. 
This is also similar to the result previously found for the AM 2306-721 pair
\citep{krabbe08}, where the companion galaxy  also presented an oxygen abundance
relatively homogeneous across the disc.

\begin{figure}
\centering
\includegraphics*[angle=-90,width=\columnwidth]{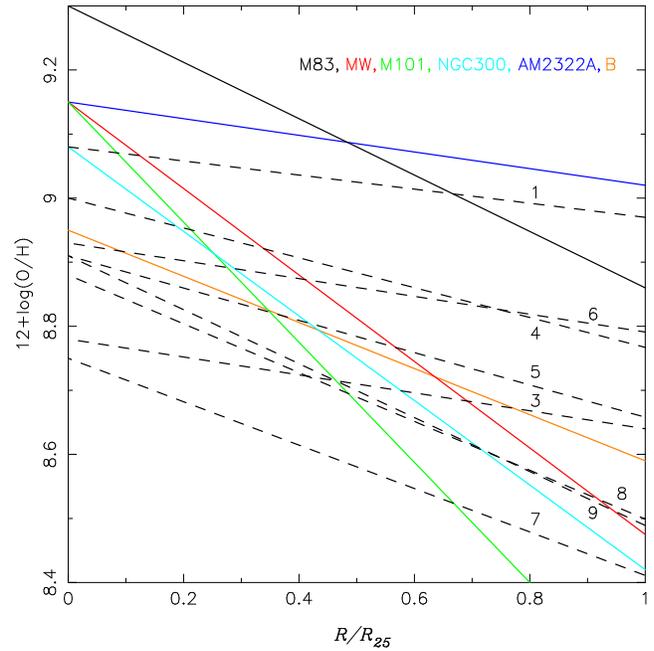}			
\caption{Metallicity gradients for AM2322A (blue line) and AM2322B (orange line).  For comparison,
we show the metallicity gradients for  the isolated galaxies M\,101 (green line), Milk Way (red line), M\,83 (filled black line), 
NGC\,300 (cyan line), and 8 interacting galaxies (dashed black lines).}
\label{kewleyfig}
\end{figure}

 We compare the metallicity gradients obtained for AM2322A and AM2322B
with the gradients for the sample of close pairs derived by \citet{kewley10}
and for the isolated galaxies M\,101, Milk Way, M\,83, and NGC\,300. The galactocentric distance is given in units
of $R/R_{25}$, where $R_{25}$ is the B-band isophote at a surface brightness
of 25 mag $\rm {arcsecond^{-2}}$. The $R_{25}$ adopted for  AM2322A and AM2322B
are 13.5 kpc and 4.2 kpc, respectively.
\citet{kewley10} computed the gradients using
the [\ion{N}{ii}]$\lambda$6584/[\ion{O}{ii}]$\lambda$3727-O/H relation
for the first three isolated galaxies
and for their sample pair. 
The metallicity gradient
in NGC\,300 was computed using the emission line
intensities of \ion{H}{ii} regions obtained by Bresolin et al. (2009)
and also using the method cited above.
These gradients are shown in Fig. \ref{kewleyfig}.
For consistency, we recalculate the gradients in
the  AM2322A and AM2322B   using the same method and  found  
 $(\frac{\rm {12+log(O/H)}}{R/R_{25}})=   -0.13(\pm0.03)$ and $(\frac{\rm {12+log(O/H)}}{R/R_{25}})=-0.36(\pm0.03)$, respectively.
As can be seen in Fig. \ref{kewleyfig}, these gradients
have about the same slope as close pairs ($\sim -0.25$)
and are  flatter than the gradients in the
isolated galaxies ($\sim -0.57$). These results are in agreement  with those found  by \citet{kewley10} 
in galaxy pairs. We interpreted that this shallower gradients can be explained by the action of inward and outward  radial flows of interstellar gas \citep{krabbe08}. In  isolated spirals, this mechanism seems to be weak or non existent.

N-body/SPH numerical simulations of equal-mass mergers predict that the radial
metallicity
gradients of the disk galaxies should flatten shortly after the first
pericenter, due to radial mixing of gas \citep{rupke2010}.
Although AM\,2322-821 is a system composed by galaxies of quite different mass
($M_{primary}/M_{secondary} \approx $ 11 from the models in Section
\ref{numsim}), our numerical simulations indicate that the current stage of the
merger is be about 90 Myr after first passage, which agrees with the scenario
proposed by the above authors.

In our previous work \citep{krabbe08}, we found that for the  AM\,2306-721 pair
only the companion galaxy presented an
oxygen abundance relatively homogeneous across the disc, while the main galaxy
showed a clear radial oxygen abundance gradient. Both systems, AM\,2306-721 and 
AM\,2322-821, are morphologically  similar. Each system is composed by a spiral
and an irregular galaxy. However, the  mass ratios estimated from the numerical
simulations for AM\,2306-721 is $M_{primary}/M_{secondary} \approx $ 2 and for 
AM\,2322-821 is about 11;  and the numerical simulations predicted that the
evolution of the  encounter was more violent and slower in AM\,2306-721. Then if
we assumed that there was an oxygen abundance gradient before the encounter in
the main galaxies in both systems, why the gradient was not destroyed or flattened by the gas
inflow in the main galaxy in AM\,2306-721? One important point is that star-formation rate in the nuclear region of the main galaxy
of AM\,2306-721 is about 10 times higher than the one derived for  AM\,2322-821.
These highest star-formation rate could have increased the heavy element content
in the central regions and thus could have produced steeper gradients, and in
these case, high gas inflow rate would be necessary to destroy or flatt the abundance
gradient in the main galaxy of  AM\,2306-721.

\section{Conclusions}
\label{final}
An observational study of the effects of the interaction in the kinematics, 
stellar population, and abundances of the 
galaxy pair AM\,2322-821 was conducted. The data consist  of long-slit spectra in
the wavelength range of 
3\,350 to7\,130\AA\, obtained with the Gemini Multi-Object Spetrograph at Gemini
South.
The main findings are the following:

\begin{enumerate}
\item
A fairly symmetric rotation curve for the companion (AM\,2322B) galaxy with a
deprojected velocity 
amplitude of  110 km s$^{-1}$ is obtained,  and a dynamical mass
of $ 1.1 - 1.3 \times 10^{10} M_{\sun}$ within  a radius of 4 kpc can be 
estimated using 
this deprojected velocity.
\item
Asymmetries in the radial velocity field were detected for companion, very
likely due the interaction between the galaxies.
\item 
In order to reconstruct the history of the AM\,2322-821 system and to
predict the evolution of the encounter, we modeled the interaction
between AM\,2322A and AM\,2322B through numerical N-body/hydrodynamical
simulations. The 
orbit that best reproduces the observational properties is found
to be hyperbolic, with an eccentricity $e=3.1$ and perigalacticum of
$q=10.5$ kpc; the current stage of the system would be about 90 Myr after
perigalacticum. 
\item
The companion galaxy is  dominated by a very young (t $ \leq 1\times10^{8}$ yr)
population, with the fraction of this population  
to the total flux  at $\lambda\, 5\,870\, \AA$, increasing outwards in the
galaxy disk.  
\item 
The stellar population of AM\,2322A
is heterogenous along  of the slit positions observed.
\item
The oxygen abundance spatial profiles obtained for both galaxies are relatively homogeneous across the galaxy discs.
The absence of an abundance gradient in these galaxies is interpreted as it having been  destroyed  by interaction-induced 
gas flows from the outer
parts to the centre of the galaxy.
\end{enumerate}

\section*{Acknowledgments}  
Based on observations obtained at the Gemini Observatory, which is operated by 
the Association of Universities for Research in Astronomy, Inc., under a 
cooperative agreement with the NSF on behalf of the Gemini partnership: the 
National Science Foundation (United States), the Science and Technology
Facilities 
Council (United Kingdom), the National Research Council (Canada), CONICYT
(Chile), 
the Australian Research Council (Australia), 
Minist\'erio da Ciencia e Tecnologia (Brazil), and SECYT (Argentina).

The authors would like to thank Volker Springel for providing them
with GADGET-2. I.Rogrigues also acknowledges Instituto Nacional de Pesquisas
Espaciais - INPE/MCT, Brazil, for providing computer time in one of its clusters
to run the simulations presented here. A. C. Krabbe  thanks the support of FAPESP, process  2010/01490-3.q

We also thank Ms. Alene Alder-Rangel for editing the English in this manuscript.

\end{document}